\documentclass[twocolumn,aps,showpacs,amsmath,superscriptaddress]{revtex4-1}

\usepackage{bm}
\usepackage{amsmath}
\usepackage{amssymb}
\usepackage[usenames,dvipsnames]{color}
\usepackage{graphicx}
\usepackage{dcolumn}
\usepackage{simplewick}
\usepackage{hyperref}

\begin{document}

\title{Perturbed Coupled-Cluster theory to calculate dipole polarizabilities
       of closed shell systems: Application to Ar, Kr, Xe and Rn}
\author{S. Chattopadhyay}
\affiliation{Physical Research Laboratory,
             Ahmedabad - 380009, Gujarat, 
             India}
\author{B. K. Mani}
\affiliation{Department of Physics, University of South Florida, Tampa,
             Florida 33620, USA}
\author{D. Angom}
\affiliation{Physical Research Laboratory,
             Ahmedabad - 380009, Gujarat,
             India}

\begin{abstract}
   We use perturbed relativistic coupled-cluster (PRCC) theory to calculate 
   the electric dipole polarizability of noble gas atoms Ar, Kr, Xe and Rn. 
   We also provide a detailed description of the nonlinear terms in the PRCC
   theory and consider the Dirac-Coulomb-Breit atomic Hamiltonian for the 
   calculations. We find that the largest contribution from Breit interaction 
   to the electric dipole polarizability is 0.1\%, in the case of Rn. 
   As we go from Ar to Rn, based on the pattern in the random phase
   approximation effects, the contraction of the outermost $p_{1/2}$
   due to relativistic corrections is discernible without any ambiguity. 
\end{abstract}



\maketitle


\section{Introduction}
  The knowledge of electric dipole polarizability, $\alpha$, of atoms and 
ions is required in different areas of physics and chemistry. It is the 
lowest order linear response property and relevant to a wide range of physical 
phenomena ranging from the microscopic to the macroscopic properties. To 
mention a few macroscopic properties, the dielectric constant and refractive 
index of gas are among the important ones. In the case of microscopic 
properties, the parity non-conservation in atoms \cite{khriplovich-91}, 
optical atomic clocks \cite{udem-02,diddams-04} and physics with the 
condensates of dilute atomic gases \cite{anderson-95, bradley-95, davis-95} are of current interest. For accurate theoretical calculation of $\alpha$, a
precise treatment of the electron correlation effects is very important. In the past, a wide range of atomic many body theories were used to calculate 
$\alpha$. The recent review by mitroy et al \cite{mitroy-10} gives a detailed 
overview of the atomic and ionic polarizabilities.

        In the present work we use the PRCC theory to calculate the $\alpha$ 
of the noble gas atoms. It is a theory we have developed to incorporate 
a perturbation in the conventional relativistic coupled-cluster (RCC) 
theory. In general, the coupled-cluster theory (CCT) 
\cite{coester-58, coester-60} is one of the most elegant many body theory 
which takes into account the electron correlation to all order. The details 
of the CCT and different variants are described in a recent review 
\cite{bartlett-07}. The theory has been widely used for atomic 
\cite{mani-09,nataraj-08,pal-07,geetha-01}, molecular\cite{isaev-04}, 
nuclear \cite{hagen-08} and condensed matter physics \cite{bishop-09} 
calculations. The PRCC theory is different from the previous RCC based 
theories in a number of ways. The most important one is the representation of 
the cluster operators in the PRCC theory is very different and it has the 
advantage of incorporating multiple perturbations of different ranks in the 
electronic sector. One basic technical advantage of PRCC is, it does
away with the summation over intermediate states in the first order
time-independent perturbation theory. In stead, the summation over all the
possible intermediate states within the chosen basis set is subsumed
in the perturbed cluster amplitudes. 

  For the calculations we use the no virtual pair Dirac-Coulomb-Breit 
Hamiltonian. However, to assess the importance of Breit interaction we also
carry out another series of calculations with the no virtual pair Dirac-Coulomb
Hamiltonian. We isolate the changes arising from the Breit interaction by
comparing the results from the two sets of the calculations. In contrast, till 
date, majority of the theoretical calculations of $\alpha$ have been done with 
the Dirac-Coulomb Hamiltonian. We have chosen the noble gas atoms to study 
as these systems are ideal to test the closed-shell PRCC theory. In previous 
works, $\alpha$ of the noble gas atoms were calculated in the framework of 
many body perturbation theory \cite{thakkar-92}, non-relativistic CCSDT 
\cite{soldan-01} and RCC single, double and triple (RCCSDT) excitation 
approximation \cite{nakajima-01}. In the later work, using RCCSDT,  the third 
order Douglas-Kroll method \cite{nakajima-00} was used. It is an alternative of the Foldy-Wouthuysen (FW) transformation and a quasi-relativistic treatment. 
For the single particle wave functions, we use the kinetically balanced 
Gaussian type Dirac-Hartree-Fock orbitals. The results from our PRCC theory 
calculations are in good agreement with the experimental data and consistent 
with previous calculations.

   The paper is organized as follows. In the Section. II, for completeness and 
easy reference we briefly describe the RCC theory with the Breit interaction. 
In section. III we introduce the PRCC theory and provide a detailed description of the tensor structure of the PRCC operators. In section. III B we give
the analytical structure of the PRCC equations. In section. III C we present
a diagrammatic and algebraic description of the  the non-linear terms in the 
PRCC theory. In section. IV we introduce the formal expression of the dipole 
polarizability and its representation in the PRCC theory. In the subsequent 
sections we describe the calculational part, and present the results and 
discussions. We then end with conclusions. All the results presented in this 
work and related calculations are in atomic units 
( $\hbar=m_e=e=4\pi\epsilon_0=1$). In this system of units the velocity of 
light is $\alpha ^{-1}$, the inverse of fine structure constant. For which we 
use the value of $\alpha ^{-1} = 137.035\;999\;074$ \cite{codata-10}.


\section{RCC Theory}
For the high-$Z$ atoms and ions, the Dirac-Coulomb-Breit Hamiltonian, denoted 
by $H^{\rm DCB}$, is appropriate to include the relativistic effects. 
However, there are complications associated with the negative energy 
continuum states of $H^{\rm DCB}$. These lead to variational collapse and 
{\em continuum dissolution} \cite{brown-51}. A formal approach to avoid these 
complications is to use the no-virtual-pair approximation. In this 
approximation, for a neutral atom of $N$ electrons \cite{sucher-80}
\begin{eqnarray}
   H^{\rm DCB} & = & \Lambda_{++}\sum_{i=1}^N \left [c\bm{\alpha}_i \cdot 
        \mathbf{p}_i + (\beta_i -1)c^2 - V_{N}(r_i) \right ] 
                       \nonumber \\
    & & + \sum_{i<j}\left [ \frac{1}{r_{ij}}  + g^{\rm B}(r_{ij}) \right ]
        \Lambda_{++},
\end{eqnarray}
where $\bm{\alpha}$ and $\beta$ are the Dirac matrices, $\Lambda_{++}$ is an 
operator which projects to the positive energy solutions and $V_{N}(r_{i})$ is 
the nuclear potential. Sandwiching the Hamiltonian with $\Lambda_{++}$ ensures 
that the effects of the negative energy continuum  states are neglected in the 
calculations. The last two terms, $1/r_{ij} $ and $g^{\rm B}(r_{ij})$  are the 
Coulomb and Breit interactions, respectively.  The later, Breit interaction, 
represents the inter-electron magnetic interactions and is given by
\begin{equation}
  g^{\rm B}(r_{12})= -\frac{1}{2r_{12}} \left [ \bm{\alpha}_1\cdot\bm{\alpha}_2
               + \frac{(\bm{\alpha_1}\cdot \mathbf{r}_{12})
               (\bm{\alpha_2}\cdot\mathbf{r}_{12})}{r_{12}^2}\right].
\end{equation}
The Hamiltonian satisfies the eigen-value equation
\begin{equation}
   H^{\rm DCB}|\Psi_{i}\rangle = E_{i}|\Psi_{i}\rangle , 
\end{equation}
where, $|\Psi_{i}\rangle$ is the exact atomic state and $E_i$ is the energy 
of the atomic state. In CCT the exact atomic state is given by the ansatz
\begin{equation}
|\Psi_i\rangle = e^{ T^{(0)}}|\Phi_i\rangle ,
\end{equation}
where $|\Phi_i\rangle$ is the reference state wave-function and $T^{(0)}$ is 
the unperturbed cluster operator. In case of closed-shell atom the model space 
of the ground state consist of a single Slater determinant, $|\Phi_0\rangle $ . 
For an $N$-electron closed-shell atom $T^{(0)} = \sum_{i=1}^N T_{i}^{(0)}$, 
where $i$ is the order of excitation.  In the coupled-cluster single and 
double (CCSD) excitation approximation,
\begin{equation}
T^{(0)} = T_{1}^{(0)} + T_{2}^{(0)}.
\end{equation}
The CCSD is a good starting point for structure and properties calculations 
of closed-shell atoms and ions. In the second quantized representation
\begin{subequations}
\begin{eqnarray}
T_1^{(0)} &= &\sum_{a,p} t_a^p {{a}_p^\dagger} a_a , \\
T_2^{(0)} &= &\frac{1}{2!}\sum_{a,b,p,q}t_{ab}^{pq} {{a}_p^\dagger}{{a}_q^
             \dagger}a_b a_a ,
\end{eqnarray}
\end{subequations}
where $t_{\ldots}^{\ldots}$ are cluster amplitudes, $a_i^{\dagger}$ ($a_i$)
are single particle creation (annihilation) operators and 
$abc\ldots$ ($pqr\ldots$) represent core (virtual) single particle states or
orbitals. The eigenvalue equation of the closed-shell ground state in CCT is
\begin{equation}
H^{\rm DCB} e^{{T^{(0)}}}|\Phi_i\rangle = E_0e^{{T^{(0)}}}|\Phi_i\rangle .
\end{equation}
Following similar procedure, the CC eigenvalue equation of closed-shell 
excited states may be defined as well.


\section{PRCC Theory}

 To incorporate an additional interaction Hamiltonian $H_{\rm int}$ 
perturbatively, we introduce the perturbed coupled-cluster operator 
$\mathbf{T^{(1)}}$. This means, $H_{\rm int}$ is applied once and residual 
Coulomb interaction to all order in all possible sequences. 
In general, $\mathbf{T^{(1)}}$ is a tensor operator and the multipole structure 
depends on the properties of $H_{\rm int}$. The properties and values of 
$\mathbf{T^{(1)}}$ indicate the effect of $H_{\rm int}$ to the atomic
state. With the perturbation, the modified eigenvalue equation is
\begin{equation}
(H^{\rm DCB}+\lambda H_{\rm int})|\tilde{\Psi}_i\rangle 
= \tilde{E}|\tilde{\Psi}_i\rangle ,
\end{equation}
where $\lambda$ is the perturbation parameter. Consider the case where 
$H_{\rm int}$ represents the interaction with an external static electric field 
$\mathbf{E}$. The interaction Hamiltonian is then 
$H_{\rm int} =-\sum_i\mathbf{r}_i\cdot\mathbf{E} = \mathbf{D}\cdot\mathbf{E} $,
where $\mathbf{D}$ is the many electron electric dipole operator. The
perturbed atomic state in PRCC theory is
\begin{eqnarray}
 |\tilde{\Psi}_i\rangle = e^{T^{(0)} + \lambda \mathbf{T}^{(1)}\cdot\mathbf{E}} 
 |\Phi_i\rangle = e^{T^{(0)}}\left [ 1 + \lambda \mathbf{T^{(1)}\cdot 
 \mathbf{E}} \right ] |\Phi_i\rangle .
 \label{psi_tilde}
\end{eqnarray}
This approach has the advantage of taking into account the effect of 
multiple perturbations systematically. Other than $\mathbf{E}$, $H_{\rm int}$ 
could be one of the interactions internal to the atom like Breit interaction, 
hyperfine interaction, etc. For the present work, we examine $\mathbf{T^{(1)}}$ 
arising from $\mathbf{E}$ which is parity odd and vector in the electronic 
space. 
\begin{figure}[h]
  \includegraphics[width=6cm]{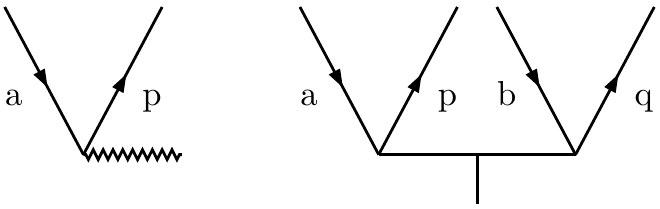}
  \caption{Diagrammatic representation of $\mathbf{T}_1^{(1)}$ and 
           $\mathbf{T}_2^{(1)} $.  }
  \label{t11t21}
\end{figure}


\subsection{Tensor structure of PRCC operator}

 For the present case, $\mathbf{E}$ as the perturbation, we can write the 
perturbed single excitation cluster operator as
\begin{equation}
  \mathbf{T}_1^{(1)} = \sum_{a,p} \tau_a^p \mathbf{C}_1 (\hat{r})
                       a_{p}^{\dagger}a_{a}.
\end{equation}  
Note that $\mathbf{T}_1^{(1)}$ is a vector operator in the electronic space 
and the $\mathbf{C}$-tensor $\mathbf{C}_1(\hat r)$ represents the vector 
nature operator. The key difference of $\mathbf{T}_1^{(1)}$ from 
$T_1^{(0)}$ is $l_a+l_p$ must be odd, in other words $(-1)^{l_a+l_p} = -1$. 
Here, $l_a$ ($l_p$) is the orbital angular momentum of the core (virtual) 
orbital $a$ ($p$). Diagrammatically, the $\mathbf{T}_1^{(1)}$
operator is represented as shown in Fig. \ref{t11t21}(a). It is similar to 
the conventional representation of $T_1^{(0)} $ but the interaction line is 
replaced by a wavy line.

 The tensor structure of $\mathbf{T}_2^{(1)} $ , on the other 
hand, has additional complications as it consists of two vertices. After 
due consideration of the $H_{\rm int}$ and $T^{(0)} $ multipole structure, it 
is represented as 
\begin{equation}
  \mathbf{T}_2^{(1)} = \sum_{a,b,p,q} \sum_{l,k} \tau_{ab}^{pq}(l,k) 
                   \{ \mathbf{C}_l(\hat{r}_1) \mathbf{C}_k(\hat{r}_2)\}^{1}
                   a_{p}^{\dagger}a_{q}^{\dagger}a_{b}a_{a}.
\end{equation}  
Like in $\mathbf{T}_1^{(1)}$,  $\mathbf{C}_k$ are the $\mathbf{C}$-tensor 
operators. Here, two $\mathbf{C}$-tensor operators of rank $l$ and $k$ are
coupled to a rank one tensor operator, $\mathbf{T}_2^{(1)}$. At the two 
vertices, the orbital angular momenta must satisfy the triangular conditions 
$|j_a - j_p| \leqslant l \leqslant (j_a + j_p)$ and 
$|j_b - j_q| \leqslant k \leqslant (j_b + j_q)$. In addition, the two tensor 
operators must  be such that $|l - k| \leqslant 1 \leqslant (l + k)$. These
selection rules arise from the triangular conditions at the vertices. The 
other selection rule follows from the parity condition. For the present
case as $H_{\rm int}$  is parity odd 
$(-1)^{(l_a + l _p)} = - (-1)^{(l_b + l _q)}$. Only then the matrix element 
$\langle pq|\mathbf{T}_2^{(1)}|ab\rangle$ is nonzero. More details on the
first principle analysis of the tensor structure, based on many-body
perturbation theory, is given in Ref. \cite{mani-11-3}. The diagrammatic
representation of $\mathbf{T}_2^{(0)}$ is as shown in Fig. \ref{t11t21}(b), 
where the vertical line on the interaction line is to represent the rank of
the operator. Further more, this representation, at a later stage, simplifies 
the angular integration using diagrams.


\subsection{PRCC equations}
The eigenvalue equation with the perturbed Hamiltonian is
\begin{equation}
  (H^{\rm DCB} + \lambda H_{\rm int}) e^{\left [T^{(0)} + \lambda 
  \mathbf{T}^{(1)}\cdot\mathbf{E} \right ]} |\Phi_0\rangle 
  = \tilde{E}_{0} e^{\left [T^{(0)} 
  + \lambda \mathbf{T}^{(1)}\cdot\mathbf{E}\right ]} |\Phi_0\rangle .
 \label{prcc_eival}
\end{equation}
When $H_{\rm int}$ is parity odd, like in the present case, there is no first
order perturbative correction to energy $\tilde{E}_0=E_0$. In the CCSD 
approximation we define the perturbed cluster operator 
$\mathbf{T}^{(1)}$ as
\begin{equation}
  \mathbf{T}^{(1)} = \mathbf{T}_1^{(1)} + \mathbf{T}_2^{(1)}.
  \label{prcc_ccsd}
\end{equation}
Using this, the PRCC equations are derived from Eq. (\ref{prcc_eival}). The 
derivation involves several operator contractions and these are more 
transparent with the normal ordered Hamiltonian
$H_{\rm N}^{\rm DCB} = H^{\rm DCB} - \langle \Phi_i|H^{\rm DCB}|\Phi_i\rangle$. 
The eigenvalue equation then assumes the form
\begin{equation}
  \left [H^{\rm DCB}_N  + \lambda H_{\rm int} \right ]|\tilde{\Psi}_0\rangle 
    = \left [ {E}_0- \langle \Phi_0|H^{\rm DCB}|\Phi_0\rangle \right ]
    |\tilde{\Psi}_0\rangle .
\end{equation}
A more convenient form of the eigenvalue equation is 
\begin{equation}
  \left (H^{\rm DCB}_{\rm N} + \lambda H_{\rm int}\right )|\tilde{\Psi}_0\rangle
   = \Delta E_0|\tilde{\Psi}_0\rangle ,
\end{equation}
where, $\Delta E_0= E_0 - \langle \Phi_0|H^{\rm DCB}|\Phi_0\rangle$ is the 
ground state correlation energy. Following the definition in 
Eq. (\ref{prcc_ccsd}), the PRCC eigen-value equation is
\begin{equation}
  \left (H^{\rm DCB}_{\rm N} + \lambda H_{\rm int}\right )e^{T^{(0)}
    + \lambda \mathbf{T}^{(1)}\cdot\mathbf{E}} |\Phi_0\rangle =
    \Delta E_0e^{T^{(0)} + \lambda \mathbf{T}^{(1)}\cdot\mathbf{E}} 
    |\Phi_0\rangle .
\end{equation}
Applying $e^{-T^{(0)}}$ from the left, we get
\begin{equation}
  \left [\bar H^{\rm DCB}_{\rm N}  + \lambda \bar H_{\rm int}\right ] 
    e^{\lambda \mathbf{T}^{(1)\cdot\mathbf{E}}}|\Phi_0\rangle 
    = \Delta E_0e^{\lambda \mathbf{T}^{(1)}\cdot\mathbf{E}} |\Phi_0\rangle ,
  \label{prcc_eq1}
\end{equation}
where $\bar H^{\rm DCB}=e^{-T^{(0)}}H^{\rm DCB}e^{T^{(0)}}$ is the similarity 
transformed Hamiltonian. Using the Campbell-Baker-Hausdorff expansion
\begin{widetext}
\begin{eqnarray}
  \bar H^{\rm DCB}  & = &  H^{\rm DCB}  +  \left [ H^{\rm DCB},T^{(0)}\right ]
   +\frac{1}{2!}\left [\left [ H^{\rm DCB},T^{(0)}\right ],T^{(0)}\right ] 
   + \frac{1}{3!}\left [\left [\left [H^{\rm DCB},T^{(0)}\right ],T^{(0)} 
   \right], T^{(0)}\right]  
                          \nonumber \\
  & & + \frac{1}{4!}\left[ \left [\left [\left [
   H^{\rm DCB},T^{(0)}\right ],T^{(0)} \right], T^{(0)}\right], T^{(0)} \right ].
\end{eqnarray}
\end{widetext}
The commutations represent contractions and as $H^{\rm DCB}$ consist of one-
and two-body interactions, the expansion terminates at the fourth order.
Multiply Eq. (\ref{prcc_eq1}) from left by $e^{-\lambda \mathbf{T}^{(1)}}$ and 
consider terms linear in $\lambda$, we get the PRCC equation
\begin{equation}
   \left [\bar{H}^{\rm DCB}_{\rm N},\mathbf{T}^{(1)}\right ]\cdot\mathbf{E} 
    + \bar{H}_{\rm int}|\Phi_0\rangle = 0 .
\end{equation}
Here, the similarity transformed interaction Hamiltonian $\bar{H}_{\rm int}$ 
terminates at second order as $H_{\rm int} $ is a one-body interaction 
Hamiltonian.

 Expanding $\bar H_{\rm N}^{\rm DCB}$ and $\bar H_{\rm int}$, the  
PRCC equation assumes the form
\begin{eqnarray}
  && \left ( \left [H_{\rm N}^{\rm DCB},\mathbf{T}^{(1)}\right ] 
     + \cdots \right ) \cdot\mathbf{E} |\Phi_0\rangle
     =  \Big ( \mathbf{D}
     \nonumber \\
  &&  + \left [\mathbf{D},T^{(0)}\right ] 
     + \frac{1}{2}\left[ \left[\mathbf{D},T^{(0)}\right ], T^{(0)}\right ] 
     \Big ) \cdot\mathbf{E}|\Phi_0\rangle .
\end{eqnarray}
Here after, for simplicity, we drop $\mathbf{E}$ from the equations and for
compact notation, we use $H_N$ to denote $H_N^{\rm DCB}$.
The cluster equations of $\mathbf{T}_1^{(1)}$ are obtained after projecting
the equation on singly excited states $\langle \Phi_a^p|$. These excitation
states, however, must be opposite in parity to $|\Phi_0 \rangle $.  
The $\mathbf{T}_2^{(1)}$ equations are obtain in a similar way after 
projecting on the doubly excited states $\langle \Phi_{ab}^{pq}|$. After the 
application of Wick's theorem, the cluster equations are
\begin{widetext}
\begin{eqnarray}
   \langle \Phi_a^p |\left [ H_N + \contraction[0.5ex]{}{H}{_{\rm N}}{T}H_{\rm N}
   \mathbf{T}^{(1)} + \contraction[0.5ex]{}{H}{_{\rm N}}{T} 
   \contraction[0.8ex]{}{H} {_{\rm N}T^{(0)}}{T} H_{\rm N}T^{(0)}
   \mathbf{T}^{(1)} + \frac{1}{2!} \contraction[0.5ex]{}{H}{_{\rm N}}{T}
   \contraction[0.8ex]{}{H}{_{\rm N}T^{(0)}}{T} 
   \contraction[1.1ex]{}{H}{_{\rm N}T^{(0)}T^{(0)}}{T} 
   H_{\rm N}T^{(0)}T^{(0)}\mathbf{T}^{(1)} \right . 
   \bigg ] |\Phi_0\rangle & = &   \langle\Phi_a^p | \left [
   \contraction[0.5ex]{}{\mathbf{D}}{}{T}\mathbf{D}T^{(0)} 
   + \frac{1}{2!} \contraction[0.5ex]{}{\mathbf{D}}{}{T}
   \contraction[0.8ex]{}{\mathbf{D}}{T^{(0)}}{T} \mathbf{D}T^{(0)}T^{(0)}
   \right ]  |\Phi_0\rangle ,  
                      \label{prcc_sing}  \\
   \langle \Phi_{ab}^{pq} |\left [ H_N + 
   \contraction[0.5ex]{}{H}{_{\rm N}}{T}H_{\rm N}
   \mathbf{T}^{(1)} + \contraction[0.5ex]{}{H}{_{\rm N}}{T} 
   \contraction[0.8ex]{}{H} {_{\rm N}T^{(0)}}{T} H_{\rm N}T^{(0)}
   \mathbf{T}^{(1)} + \frac{1}{2!} \contraction[0.5ex]{}{H}{_{\rm N}}{T}
   \contraction[0.8ex]{}{H}{_{\rm N}T^{(0)}}{T} 
   \contraction[1.1ex]{}{H}{_{\rm N}T^{(0)}T^{(0)}}{T} 
   H_{\rm N}T^{(0)}T^{(0)}\mathbf{T}^{(1)} + \cdots \right ] |\Phi_0\rangle 
   & = &  \langle\Phi_{ab}^{pq} | \left [
   \contraction[0.5ex]{}{\mathbf{D}}{}{T}\mathbf{D}T^{(0)} 
   + \frac{1}{2!} \contraction[0.5ex]{}{\mathbf{D}}{}{T}
   \contraction[0.8ex]{}{\mathbf{D}}{T^{(0)}}{T} \mathbf{D}T^{(0)}T^{(0)}
   \right ]  |\Phi_0\rangle . 
                      \label{prcc_dbl}
\end{eqnarray}
\end{widetext}
Where $\contraction[0.5ex]{}{A}{}{B}AB $ represents all possible contractions
between the two operators $A$ and $B$. The Eq. (\ref{prcc_sing}) and 
(\ref{prcc_dbl}) form a set of coupled nonlinear algebraic equations. 
However, $T^{(0)} $ are solved first as these are independent of 
$\mathbf{T}^{(1)}$, the PRCC equations are then reduced to coupled linear 
algebraic equations. An approximation which incorporates all the 
important many-body effects like random phase approximation (RPA) is the 
linearized PRCC (LPRCC). In this approximation, only the terms linear in 
$T$, equivalent to retaining only 
$\contraction[0.5ex]{}{H}{_{\rm N}}{T}H_{\rm N}T^{(1)}$ 
and $\contraction[0.5ex]{}{\mathbf{r}}{_i}{T}\mathbf{r}_iT^{(0)} $ in 
Eq. (\ref{prcc_sing}) and (\ref{prcc_dbl}), are considered in the equations. 
Here after we use $T$ as the general representation of the both $T^{(0)}$ and
$\mathbf{T}^{(1)}$ operators.


\subsection{Nonlinear terms in PRCC}
  The calculations with the LPRCC approximation involves few many-body 
diagrams, and it is computationally less complex and hence faster. In our 
calculations, the LPRCC equations are solved first and we use the solutions as 
the initial guess to solve the PRCC equations. To describe the PRCC equations 
in detail, we examine each of the nonlinear terms. To begin with consider the 
second term on the left hand side of Eq. (\ref{prcc_sing}) and 
(\ref{prcc_dbl}), second order in $T$, in CCSD approximation it expands to 
\begin{eqnarray}
  \contraction[0.5ex]{}{H}{_{\rm N}}{T}
  \contraction[0.8ex]{}{H}{_{\rm N}T^{(0)}}{\mathbf{T}} H_{\rm N}T^{(0)}
  \mathbf{T}^{(1)} & = &
  \contraction[0.5ex]{}{H}{_{\rm N}}{T}
  \contraction[0.8ex]{}{H}{_{\rm N}T^{(0)}}{\mathbf{T}} H_{\rm N}T_1^{(0)}
  \mathbf{T}_1^{(1)} + \contraction[0.5ex]{}{H}{_{\rm N}}{T}
  \contraction[0.8ex]{}{H}{_{\rm N}T^{(0)}}{\mathbf{T}} H_{\rm N}T_1^{(0)}
  \mathbf{T}_2^{(1)} 
                    \nonumber \\
& & +\contraction[0.5ex]{}{H}{_{\rm N}}{T} \contraction[0.8ex]{}{H}{_{\rm N}
    T^{(0)}}{\mathbf{T}} H_{\rm N}T_2^{(0)}\mathbf{T}_1^{(1)} +
    \contraction[0.5ex]{}{H}{_{\rm N}}{T}\contraction[0.8ex]{}{H}{_{\rm N}
    T^{(0)}}{\mathbf{T}} H_{\rm N}T_2^{(0)} \mathbf{T}_2^{(1)}. 
 \label{2nd_ord}
\end{eqnarray}
All the terms contribute to both $\mathbf{T}_1^{(1)} $ and $\mathbf{T}_2^{(1)}$.
Similarly, the third term on the left hand side of Eq. (\ref{prcc_sing}) and
(\ref{prcc_dbl}), third order in cluster amplitude, expands to expands to
\begin{eqnarray}
  \contraction[0.5ex]{}{H}{_{\rm N}}{T}
  \contraction[0.8ex]{}{H}{_{\rm N}T^{(0)}}{T} 
  \contraction[1.1ex]{}{H}{_{\rm N}T^{(0)}T^{(0)}}{\mathbf{T}} 
  H_{\rm N}T^{(0)}T^{(0)}\mathbf{T}^{(1)} & = & \contraction[0.5ex]{}{H}
  {_{\rm N}}{T} \contraction[0.8ex]{}{H}{_{\rm N}T^{(0)}}{T} 
  \contraction[1.1ex]{}{H}{_{\rm N}T^{(0)}T^{(0)}}{\mathbf{T}} 
  H_{\rm N}T_1^{(0)}T_1^{(0)}\mathbf{T}_1^{(1)}  
  + \contraction[0.5ex]{}{H}{_{\rm N}}{T}
  \contraction[0.8ex]{}{H}{_{\rm N}T^{(0)}}{T} 
  \contraction[1.1ex]{}{H}{_{\rm N}T^{(0)}T^{(0)}}{\mathbf{T}} 
  H_{\rm N}T_1^{(0)}T_2^{(0)}\mathbf{T}_1^{(1)} 
              \nonumber \\ 
 &&\!\!\!\!\!\!\!\!\!\!\!\!\!\!\!\! + \contraction[0.5ex]{}{H}{_{\rm N}}{T}
  \contraction[0.8ex]{}{H}{_{\rm N}T^{(0)}}{T} 
  \contraction[1.1ex]{}{H}{_{\rm N}T^{(0)}T^{(0)}}{\mathbf{T}} 
  H_{\rm N}T_1^{(0)}T_1^{(0)}\mathbf{T}_2^{(1)} 
  + \contraction[0.5ex]{}{H}{_{\rm N}}{T}
  \contraction[0.8ex]{}{H}{_{\rm N}T^{(0)}}{T} 
  \contraction[1.1ex]{}{H}{_{\rm N}T^{(0)}T^{(0)}}{\mathbf{T}} 
  H_{\rm N}T_1^{(0)}T_2^{(0)}\mathbf{T}_2^{(1)} .
 \label{3rd_ord}
\end{eqnarray}
In this equation, out of the four terms, only the first one contributes to 
$\mathbf{T}_1^{(1)}$. But, all the terms contribute to the $\mathbf{T}_2^{(1)}$.
At the fourth order there is only one term and it contributes to only 
$\mathbf{T}_2^{(1)}$. The terms on the right hand side of Eq. (\ref{prcc_sing}) 
and (\ref{prcc_dbl}) expands to
\begin{eqnarray}
 \contraction[0.5ex]{}{\mathbf{D}}{}{T}\mathbf{D}T^{(0)}  & = &
 \contraction[0.5ex]{}{\mathbf{D}}{}{T}\mathbf{D}T_1^{(0)}  + 
 \contraction[0.5ex]{}{\mathbf{D}}{}{T}\mathbf{D}T_2^{(0)} ,  \\
 \contraction[0.5ex]{}{\mathbf{D}}{}{T}
 \contraction[0.8ex]{}{\mathbf{D}}{T^{(0)}}{T} \mathbf{D}T^{(0)}T^{(0)} &= &
 \contraction[0.5ex]{}{\mathbf{D}}{}{T}
 \contraction[0.8ex]{}{\mathbf{D}}{T^{(0)}}{T} \mathbf{D}T_1^{(0)}T_1^{(0)}
 + \contraction[0.5ex]{}{\mathbf{D}}{}{T}
   \contraction[0.8ex]{}{\mathbf{D}}{T^{(0)}}{T}\mathbf{D}T_1^{(0)}T_2^{(0)}.
\end{eqnarray}
Here, $\contraction[0.5ex]{}{\mathbf{D}}{}{T}\mathbf{D}T_1^{(0)}  $ 
and $\contraction[0.5ex]{}{\mathbf{D}}{}{T}
\contraction[0.8ex]{}{\mathbf{D}}{T^{(0)}}{T} \mathbf{D}T_1^{(0)}T_2^{(0)}$ 
are nonzero only for $\mathbf{T}_1^{(1)} $ and 
$\mathbf{T}_2^{(1)} $, respectively. Each of the terms, after contraction,
generate several topologically unique Goldstone diagrams. At this stage, the 
diagrammatic treatment is the preferred mode of further analysis and 
calculation. It simplifies the calculations and is well suited to represent 
contractions between the operators. 
\begin{figure}[h]
  \includegraphics[width=8cm]{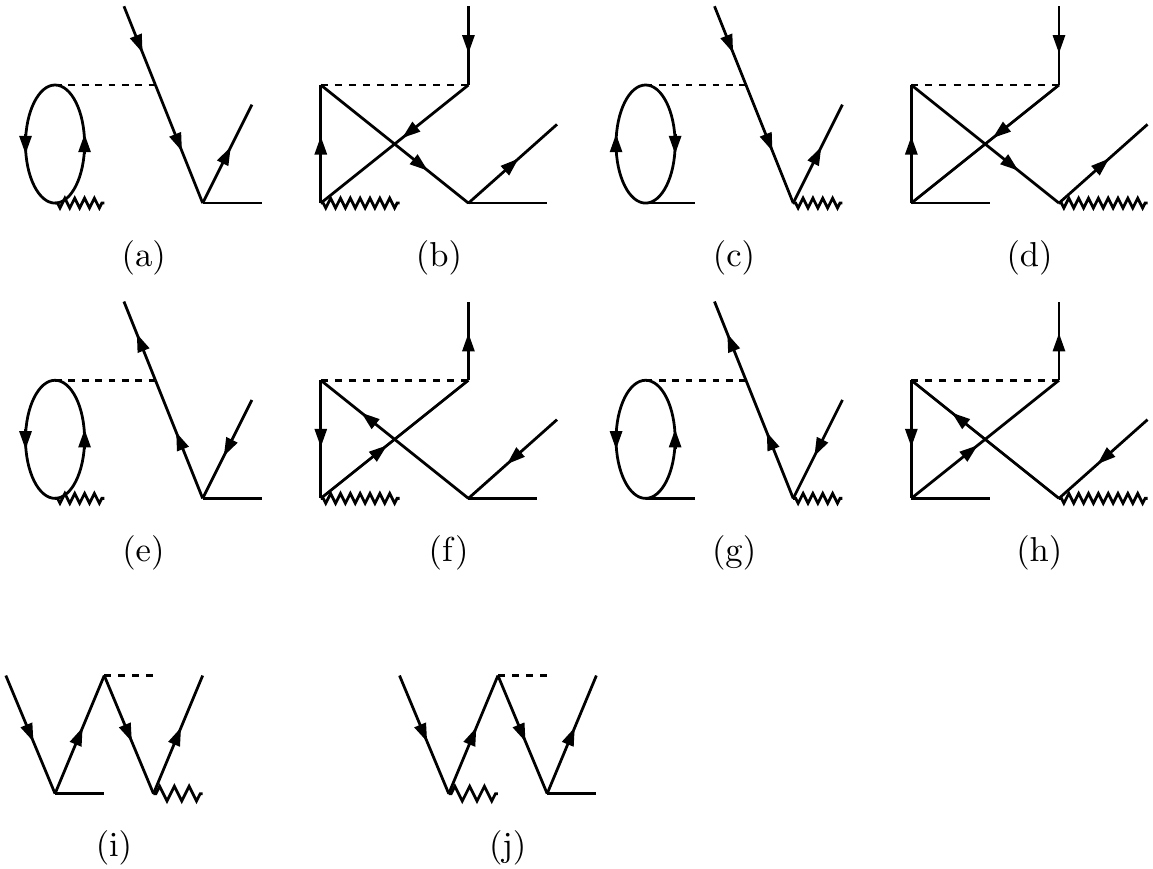}
  \caption{Diagrams of $\mathbf{T}_1^{(1)}$ arising from 
   $\contraction[0.5ex]{}{H}{_{\rm N}}{T}
    \contraction[0.8ex]{}{H}{_{\rm N}T^{(0)}}{T} 
     H_{\rm N}T_1^{(0)}\mathbf{T}_1^{(1)} $}.
  \label{ht10t11}
\end{figure}


\subsubsection{$\mathbf{T}_1^{(1)} $ diagrams}
\begin{figure}[h]
 \includegraphics[width=8cm]{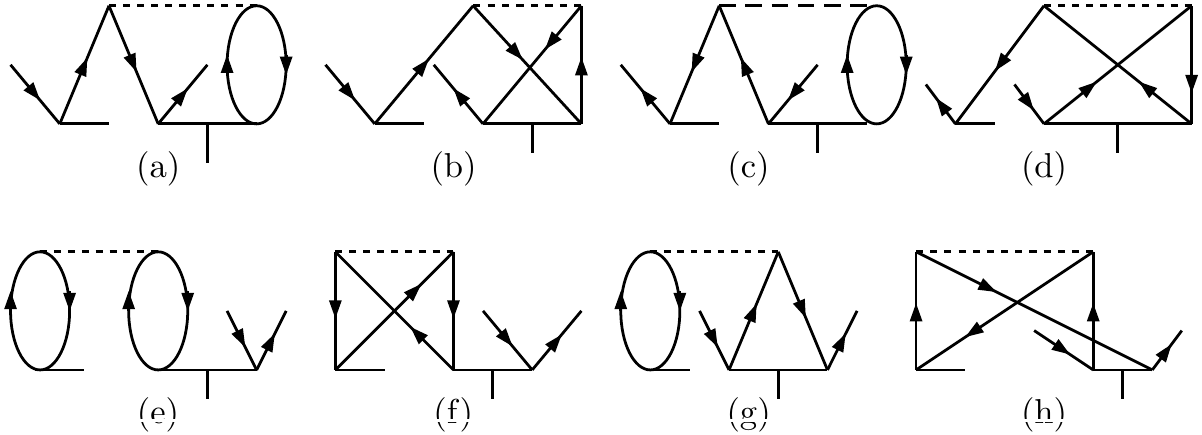}
 \caption{Diagrams arising from the contraction
 $\contraction[0.5ex]{}{H}{_{\rm N}}{T}
  \contraction[0.8ex]{}{H}{_{\rm N}T^{(0)}}{T} 
  H_{\rm N}T_1^{(0)}\mathbf{T}_2^{(1)}$} 
 \label{st10t21}
\end{figure}

\begin{figure}[h]
 \includegraphics[width=8cm]{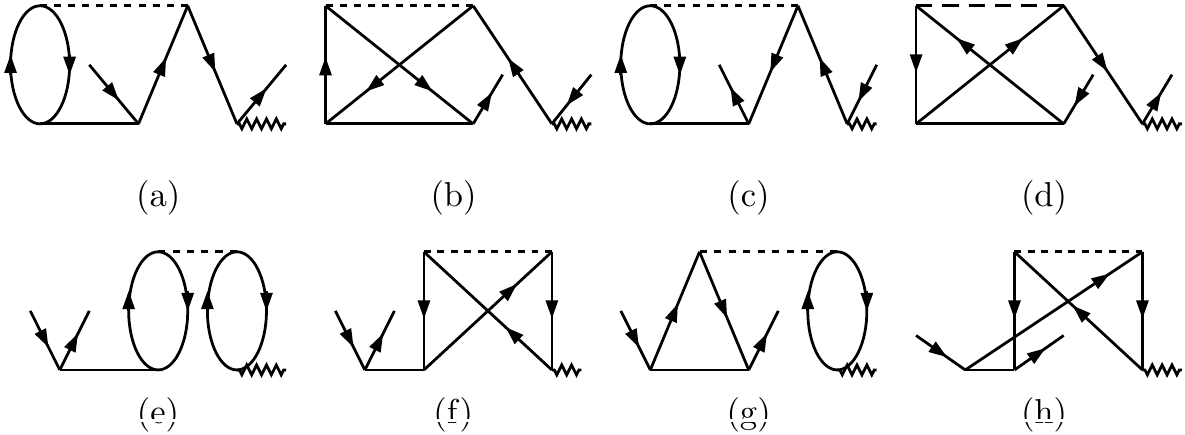}
 \caption{Diagrams arising from the contraction
 $\contraction[0.5ex]{}{H}{_{\rm N}}{T}
  \contraction[0.8ex]{}{H}{_{\rm N}T^{(0)}}{T} 
  H_{\rm N}T_2^{(0)}\mathbf{T}_1^{(1)}$}
 \label{st21t10}
\end{figure}

In this section we describe the single excitation diagrams arising from the  
non linear terms. The  many-body diagrams or the Goldstone diagrams are
drawn and evaluated as described in ref. \cite{lindgren-86}. The evaluation
of the diagrams consists of calculating the radial and angular integrals. 
Consider the first term on the right hand side of Eq. (\ref{2nd_ord}),
$ \contraction[0.5ex]{}{H}{_{\rm N}}{T}\contraction[0.8ex]{}{H}{_{\rm N}
T^{(0)}}{T} H_{\rm N}T_1^{(0)} \mathbf{T}_1^{(1)}$, it is 
equivalent to ten diagrams and these are shown in Fig. \ref{ht10t11}.
Algebraically, we can write it as
\begin{eqnarray}
  \langle \contraction[0.5ex]{}{H}{_{\rm N}}{T}
   \contraction[0.8ex]{}{H}{_{\rm N}T^{(0)}}{T} H_{\rm N}T_1^{(0)}
   \mathbf{T}_1^{(1)}\rangle_a^p & = & \sum_{bcqa} \tilde{g}_{bcqa}\left ( 
   t_{c}^{p}\tau_{b}^{q} + t_{b}^{q}\tau_{c}^{p} \right ) 
       \nonumber \\
&& + \sum_{bpqr} \tilde{g}_{bpqr} \left ( t_{a}^{r}\tau_{b}^{q} 
   + t_{b}^{q}\tau_{a}^{r}  \right ), \nonumber
\end{eqnarray}
where $g_{ijkl} = \langle ij|1/r_{12} + g^{\rm B}(r_{12})|kl\rangle$ is the 
matrix element of the electron-electron interactions and 
$\tilde{g}_{ijkl}=g_{ijkl}-g_{ijlk}$ is the antisymmtrized matrix element. We
have used $\langle \cdots\rangle_a^p $ to represent the matrix element
$\langle\Phi_a^p| \cdots  |\Phi_0\rangle$. The 
diagrams in Fig. \ref{ht10t11}(i-j), arising from the one-body part of 
$H_N$, evaluate to zero when orbitals are calculated with 
Dirac-Hartree-Fock potential. The next term,
$\contraction[0.5ex]{}{H}{_{\rm N}}{T}\contraction[0.8ex]{}{H}{_{\rm N}
T^{(0)}}{T} H_{\rm N}T_1^{(0)}\mathbf{T}_2^{(1)}$, generates eight 
diagrams and these are shown in Fig. \ref{st10t21}.
It is to be noted that contractions with only the $g_{abpq}$ type of two-body 
interaction are non-zero. The algebraic expression of the diagrams is
\begin{eqnarray}
  \langle \contraction[0.5ex]{}{H}{_{\rm N}}{T}\contraction[0.8ex]{}{H}{_{\rm N}
  T^{(0)}}{T} H_{\rm N}T_1^{(0)}\mathbf{T}_2^{(1)}\rangle_a^p & = &
  \sum_{bcqr} \tilde{g}_{cbrq} \left ( t_{a}^{r}\tau_{pq}^{cb} + 
   t_{c}^{p}\tau_{ab}^{rq} + t_{c}^{r}\tau_{ba}^{qp}\right .
           \nonumber \\
  && + \left.  t_{b}^{q}\tau_{ac}^{rp} \right ).  \nonumber
\end{eqnarray}
At the second order $\contraction[0.5ex]{}{H}{_{\rm N}}{T}
\contraction[0.8ex]{}{H}{_{\rm N}T^{(0)}}{T} H_{\rm N}T_2^{(0)}
\mathbf{T}_1^{(1)}$ is the last term. Like the previous term, after contraction
it generates eight diagrams and these are shown in Fig. \ref{st21t10}. 
The topological structure of the diagrams are very similar to those of 
Fig. \ref{st10t21}. The algebraic expression of the diagrams is
\begin{eqnarray}
  \langle \contraction[0.5ex]{}{H}{_{\rm N}}{T} 
  \contraction[0.8ex]{}{H}{_{\rm N}T^{(0)}}{T} H_{\rm N}T_2^{(0)}
  \mathbf{T}_1^{(1)}\rangle_a^p &= & \sum_{bcqr} \tilde{g}_{bcqr} \left ( 
  t_{ba}^{qr}\tau_{p}^{c} + t_{bc}^{qp}\tau_{a}^{r} + t_{ab}^{pq}\tau_{c}^{r} 
  \right .              \nonumber \\
  && + \left. t_{ac}^{rp}\tau_{b}^{q} \right ) . \nonumber 
\end{eqnarray}

\begin{figure}[h]
 \includegraphics[width=8cm]{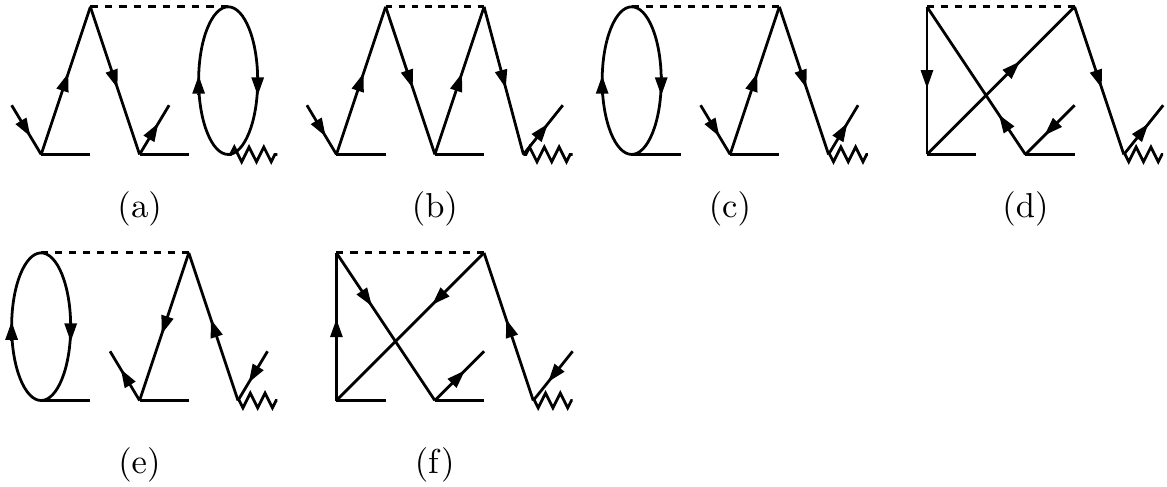}
 \caption{Diagrams arising from the contraction
 $\contraction[0.5ex]{}{H}{_{\rm N}}{T}
  \contraction[0.8ex]{}{H}{_{\rm N}T^{(0)}}{T} 
  \contraction[1.1ex]{}{H}{_{\rm N}T^{(0)}T^{(0)}}{T} 
  H_{\rm N}T_1^{(0)}T_1^{(0)}\mathbf{T}_1^{(1)} $}
 \label{st10t10t11}
\end{figure}

At the third order, as mentioned earlier, only 
$\contraction[0.5ex]{}{H}{_{\rm N}}{T}
\contraction[0.8ex]{}{H}{_{\rm N}T^{(0)}}{T} 
\contraction[1.1ex]{}{H}{_{\rm N}T^{(0)}T^{(0)}}{T}
H_{\rm N}T_1^{(0)}T_1^{(0)}\mathbf{T}_1^{(1)} $ contributes to the
$\mathbf{T}_1^{(1)}$ diagrams. This term generate six Goldstone diagrams and 
these are shown in Fig. \ref{st10t10t11}. The 
algebraic expression of the diagrams is
\begin{eqnarray}
  \langle\contraction[0.5ex]{}{H}{_{\rm N}}{T}
  \contraction[0.8ex]{}{H}{_{\rm N}T^{(0)}}{T} 
  \contraction[1.1ex]{}{H}{_{\rm N}T^{(0)}T^{(0)}}{T}
  H_{\rm N}T_1^{(0)}T_1^{(0)}\mathbf{T}_1^{(1)}\rangle_a^p & = &
  \sum_{bcqr} \tilde{g}_{bcqr} \left ( t_{a}^{r}t_{c}^{p}\tau_{q}^{b} + 
   t_{b}^{q}t_{a}^{r}\tau_{c}^{p} \right .
           \nonumber \\
 && + \left . t_{b}^{q}t_{c}^{p}\tau_{a}^{r} \right ).  \nonumber
\end{eqnarray}
In total, the nonlinear terms in the $\mathbf{T}_1^{(1)}$  cluster equation
generate thirty Goldstone diagrams. Considering that $T_2^{(0)}$  and 
$\mathbf{T}_1^{(1)}$ are the dominant cluster operators in the unperturbed RCC 
and PRCC, respectively, we can expect the magnitude of 
$\contraction[0.5ex]{}{H}{_{\rm N}}{T}\contraction[0.8ex]{}{H}{_{\rm N}
T^{(0)}}{T} H_{\rm N}T_2^{(0)} \mathbf{T}_1^{(1)}$ to be largest.


\subsubsection{$\mathbf{T}_2^{(1)}$ diagrams}
\begin{figure}[h]
 \includegraphics[width=8cm]{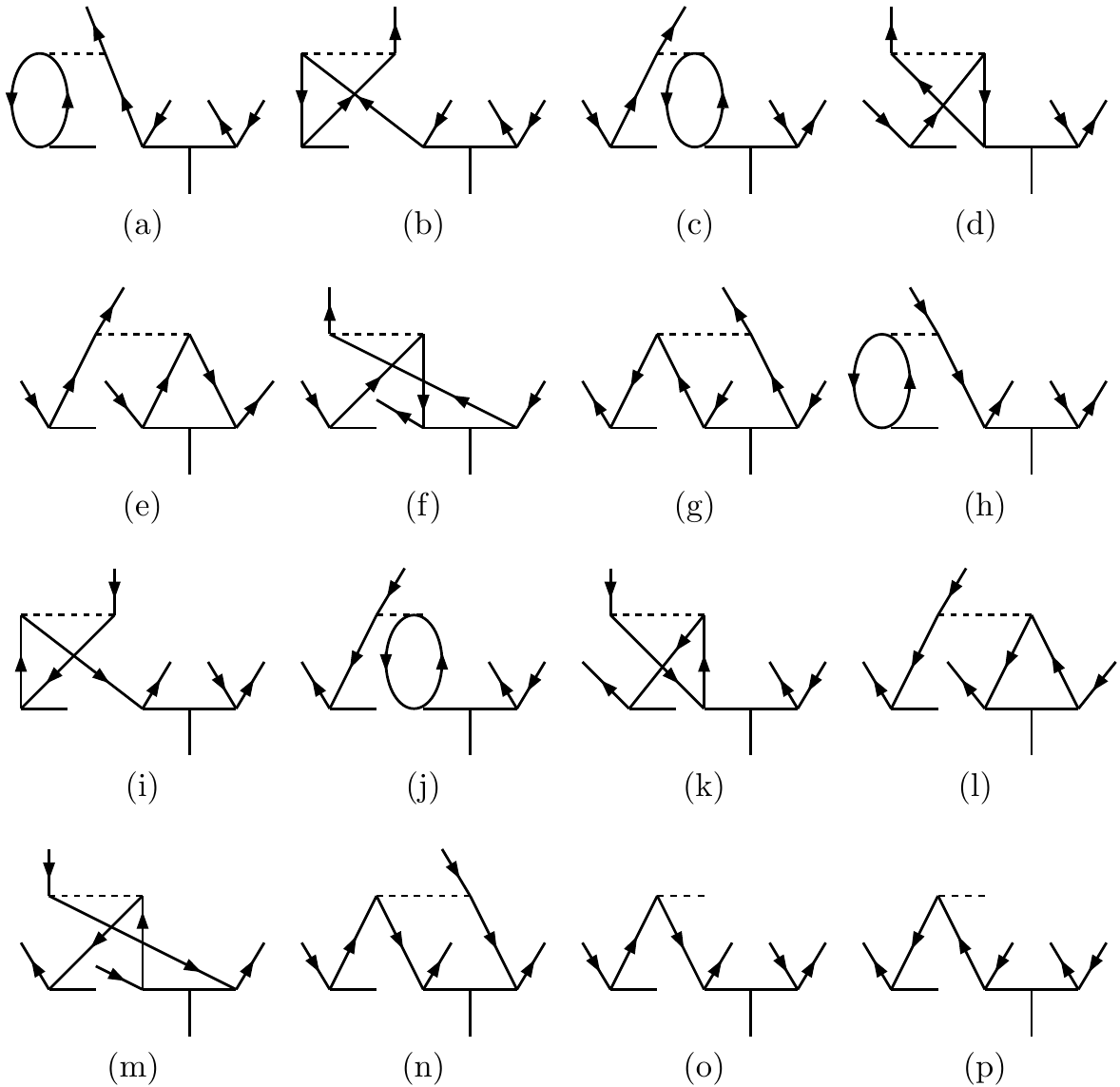}
 \caption{Diagrams arising from the contraction
 $\contraction[0.5ex]{}{H}{_{\rm N}}{T}
  \contraction[0.8ex]{}{H}{_{\rm N}T^{(0)}}{T} 
  H_{\rm N}T_1^{(0)}\mathbf{T}_2^{(1)}$}
 \label{dt10t21}
\end{figure}

\begin{figure}[h]
 \includegraphics[width=8cm]{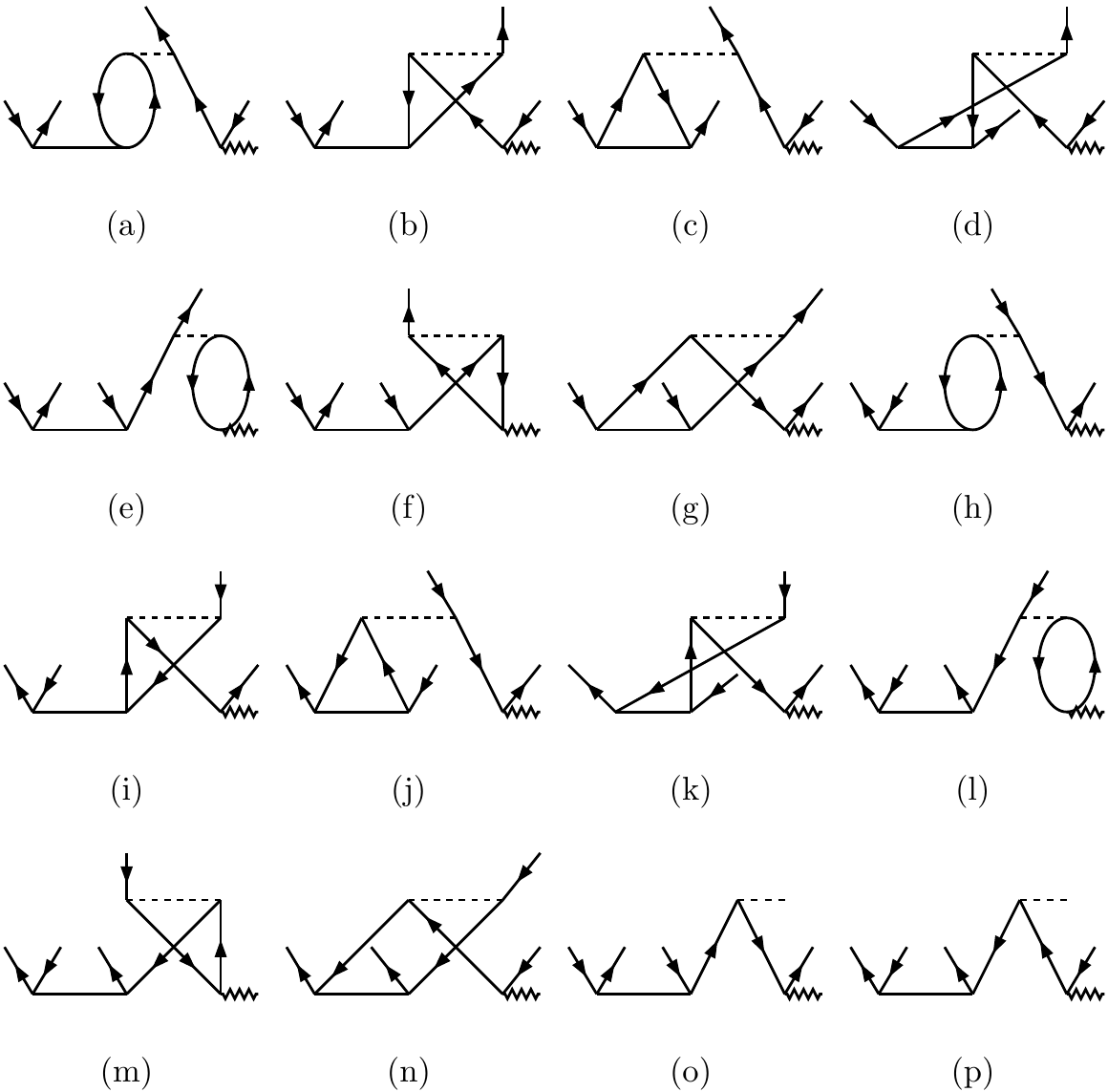}
 \caption{Diagrams arising from the contraction
 $\contraction[0.5ex]{}{H}{_{\rm N}}{T}
  \contraction[0.8ex]{}{H}{_{\rm N}T^{(0)}}{T} 
  H_{\rm N}T_2^{(0)}\mathbf{T}_1^{(1)}$}
 \label{dt20t11}
\end{figure}

\begin{figure}[h]
 \includegraphics[width=8cm]{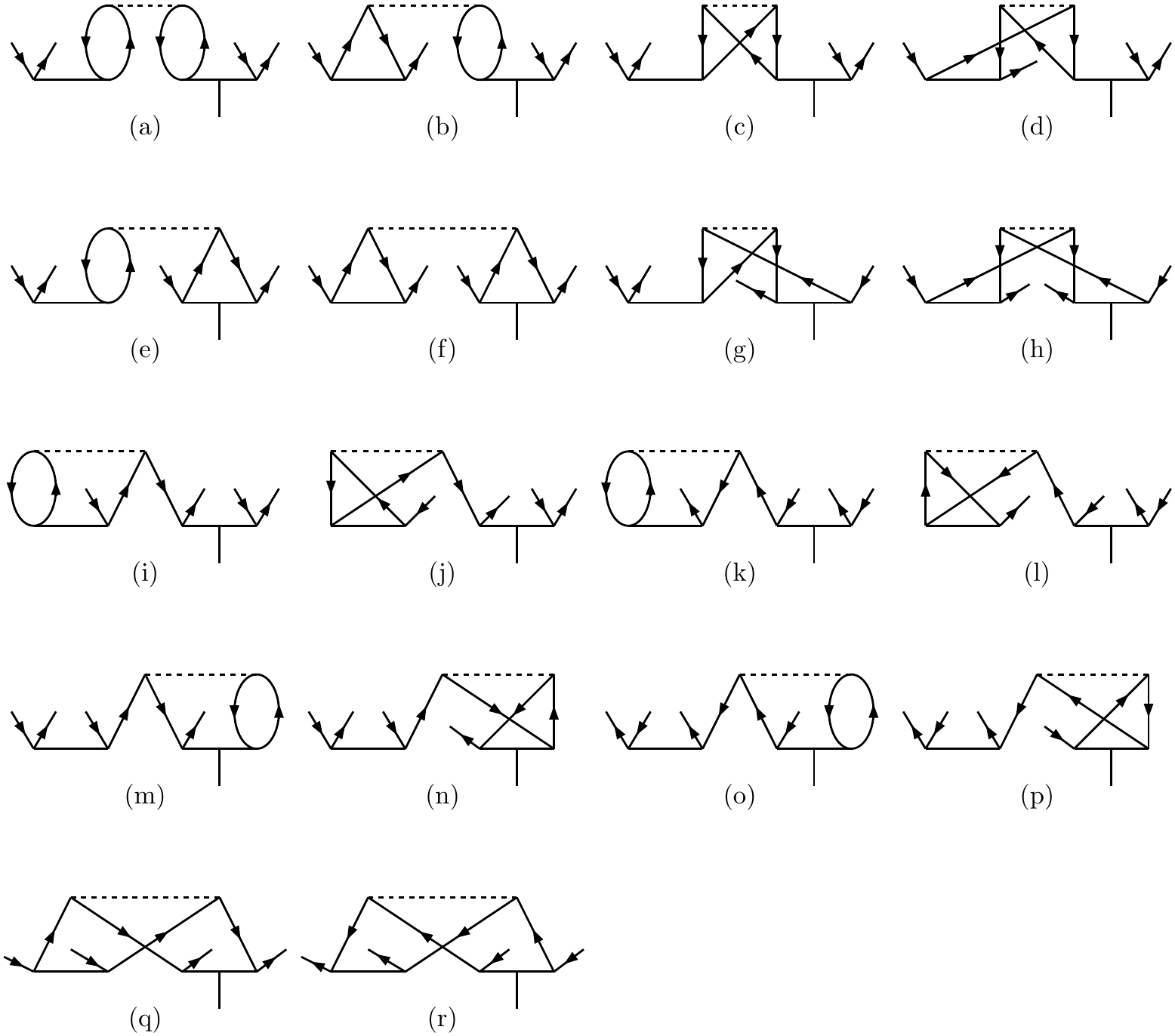}
 \caption{Diagrams arising from the contraction
  $\contraction[0.5ex]{}{H}{_{\rm N}}{T}
  \contraction[0.8ex]{}{H}{_{\rm N}T^{(0)}}{T} 
  H_{\rm N}T_2^{(0)}\mathbf{T}_2^{(1)}$}
 \label{dt20t21}
\end{figure}

In this section we discuss the Goldstone diagrams of $\mathbf{T}_2^{(1)}$ 
arising from the non-linear terms on the left hand side of Eq. (\ref{prcc_dbl}). 
Consider the second order term, after expansion there are four terms as given
in Eq. (\ref{2nd_ord}) and all have nonzero contribution to $\mathbf{T}_2^{(1)}$.
\begin{figure}[h]
 \includegraphics[width=8cm]{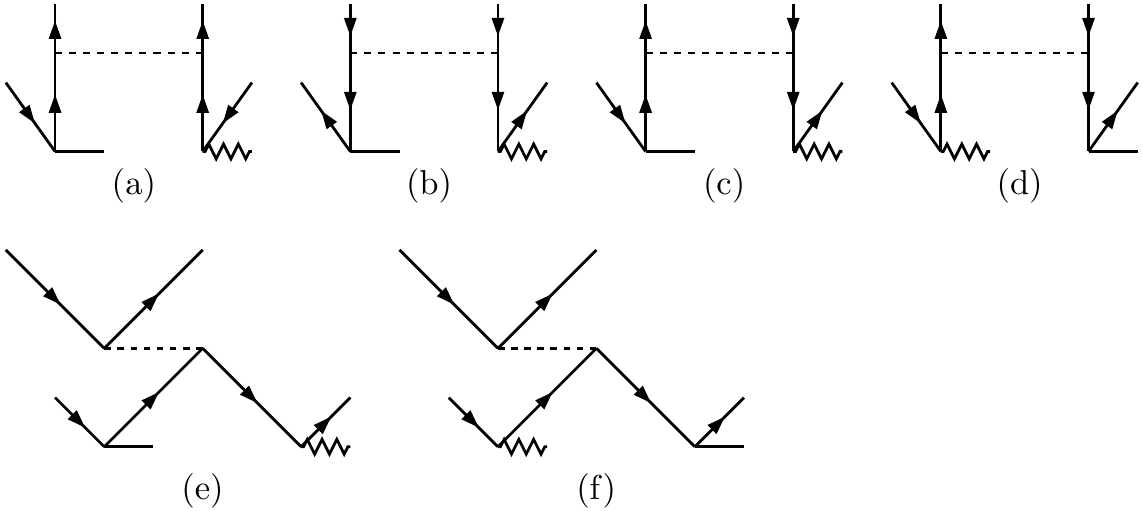}
 \caption{Diagrams arising from the contraction
 $\contraction[0.5ex]{}{H}{_{\rm N}}{T}
  \contraction[0.8ex]{}{H}{_{\rm N}T^{(0)}}{T} 
  H_{\rm N}T_1^{(0)}\mathbf{T}_1^{(1)} $}
 \label{dt10t11}
\end{figure}
The first term,  $\contraction[0.5ex]{}{H}{_{\rm N}}{T}
\contraction[0.8ex]{}{H}{_{\rm N}T^{(0)}}{T} 
H_{\rm N}T_1^{(0)}\mathbf{T}_1^{(1)} $, has six diagrams and these are shown
in Fig. \ref{dt10t11}. The equivalent algebraic expression is 
\begin{eqnarray}
  \langle \contraction[0.5ex]{}{H}{_{\rm N}}{T}
  \contraction[0.8ex]{}{H}{_{\rm N}T^{(0)}}{T} 
  H_{\rm N}T_1^{(0)}\mathbf{T}_1^{(1)}\rangle_{ab}^{pq} & = & \sum_{rs} g_{pqrs}
  t_{a}^{r} \tau_{b}^{s} + \sum_{cd} g_{cdab}t_{c}^{p}\tau_{d}^{q} 
  - \sum_{cr} g_{pcrb} \nonumber \\ 
 & & \times \left [  (t_a^r + t_b^r)\tau_c^q + t_c^q (\tau_a^r
  + \tau_b^r) \right ],  \nonumber
\end{eqnarray}
where, we have used $\langle\cdots\rangle_{ab}^{pq}$ to represent the matrix
element $\langle\Phi_{ab}^{pq}| \cdots |\Phi_0\rangle$. The next term, 
$\contraction[0.5ex]{}{H}{_{\rm N}}{T}
  \contraction[0.8ex]{}{H}{_{\rm N}T^{(0)}}{T} 
  H_{\rm N}T_1^{(0)}\mathbf{T}_2^{(1)}$, has sixteen diagrams and these
are shown in Fig. \ref{dt10t21}.
However, the last two diagrams are zero when Dirac-Hartree-Fock-Breit orbitals 
are used, like we do in the present work, in the PRCC calculations. The 
equivalent algebraic expression is 
\begin{eqnarray}
  \langle \contraction[0.5ex]{}{H}{_{\rm N}}{T}
  \contraction[0.8ex]{}{H}{_{\rm N}T^{(0)}}{T} 
  H_{\rm N}T_1^{(0)}\mathbf{T}_2^{(1)}\rangle_{ab}^{pq} & = & \sum_{crs} g_{cqrs} 
  \left ( t_{c}^{r}\tau_{ba}^{sp} 
   - t_{c}^{s}\tau_{ba}^{rp} + t_{b}^{s}\tilde{\tau}_{ca}^{rp}
   - t_{b}^{r}\tau_{ca}^{sp} \right . 
                   \nonumber \\
 & & \left . - t_{a}^{r}\tau_{cb}^{ps} - t_{c}^{p}\tau_{ab}^{rs} \right ) 
   + \sum_{cdr} g_{cdrb} \left ( -t_{c}^{r}\tau_{da}^{qp}  \right .
                   \nonumber \\
 & & \left . + t_{d}^{r}\tau_{ca}^{qp} - t_{d}^{q}\tilde{\tau}_{ca}^{rp}
   + t_{c}^{p}\tau_{ad}^{rq} + t_{a}^{r}\tau_{cd}^{pq} \right ), \nonumber
\end{eqnarray}
where, $\tilde{\tau}_{ca}^{rp} = \tau_{ca}^{rp}-\tau_{ac}^{rp}$ is the 
antisymmetrised amplitude of $\mathbf{T}_2^{(1)} $ . Interchanging the
order of excitations of the cluster operators, we get the next term 
$\contraction[0.5ex]{}{H}{_{\rm N}}{T}
\contraction[0.8ex]{}{H}{_{\rm N}T^{(0)}}{T} 
H_{\rm N}T_2^{(0)}\mathbf{T}_1^{(1)}$. Like in the previous term there are
sixteen diagrams and these are shown in Fig. \ref{dt20t11} and
equivalent algebraic expression is
\begin{eqnarray}
   \langle \contraction[0.5ex]{}{H}{_{\rm N}}{T}
   \contraction[0.8ex]{}{H}{_{\rm N}T^{(0)}}{T} 
   H_{\rm N}T_2^{(0)}\mathbf{T}_1^{(1)}\rangle_{ab}^{pq} & = & \sum_{crs} 
   g_{cqrs} \left ( \tilde{t}_{ac}^{pr}\tau_{b}^{s} - t_{ac}^{ps}\tau_{b}^{r} 
    - t_{bc}^{sp}\tau_{a}^{r} + t_{ab}^{ps}\tau_{c}^{r} \right .
                  \nonumber \\
  & & \left . - t_{ab}^{pr}\tau_{c}^{s} - t_{ab}^{rs}\tau_{c}^{p} \right )
    + \sum_{cdr} g_{cdrb} \left ( \tilde{t}_{ca}^{pr}\tau_{d}^{q} \right.
                  \nonumber \\
  & & - t_{ad}^{pr}\tau_{c}^{q} + t_{da}^{qr}\tau_{c}^{p}
      - t_{ad}^{pq}\tau_{c}^{r} + t_{ac}^{pq}\tau_{d}^{r} 
                  \nonumber \\
  & & \left . + t_{cd}^{pq}\tau_{a}^{r} \right ),   \nonumber
\end{eqnarray}
where, $\tilde{t}_{ac}^{pr}=t_{ac}^{pr} -t_{ac}^{rp} $ is the antisymmetrised amplitude of $T_2^{(0)}$.  
The last second order term is $\contraction[0.5ex]{}{H}{_{\rm N}}{T}
  \contraction[0.8ex]{}{H}{_{\rm N}T^{(0)}}{T} 
  H_{\rm N}T_2^{(0)}\mathbf{T}_2^{(1)}$ and we can expect a large number of 
diagrams as both of the cluster operators are double excitation. There are
sixteen diagrams and these are shown in Fig. \ref{dt20t21}. 
The algebraic expression for the diagrams is 
\begin{eqnarray}
 \!\!\!\!\!\!\!\!\! \langle \contraction[0.5ex]{}{H}{_{\rm N}}{T}
  \contraction[0.8ex]{}{H}{_{\rm N}T^{(0)}}{T} 
  H_{\rm N}T_2^{(0)}\mathbf{T}_2^{(1)} \rangle_{ab}^{pq} & = &
  \sum_{cdrs} g_{cdrs} \left ( \tilde{t}_{ac}^{pr}\tilde{\tau}_{db}^{sq} 
  - \tilde{t}_{ac}^{ps}\tau_{db}^{rq} + t_{ac}^{ps}\tau_{db}^{qr} \right. 
                        \nonumber \\
  & & + t_{ac}^{sq}\tau_{db}^{pr} - \tilde{t}_{ca}^{rs}\tau_{db}^{pq} 
   - \tilde{t}_{cd}^{rp}\tau_{ab}^{sq}- t_{ab}^{ps}\tau_{dc}^{qr}  
                        \nonumber \\
  & & \left. + t_{ab}^{pr}\tau_{dc}^{qs} - t_{ac}^{pq}\tilde{\tau}_{bd}^{rs} 
   + t_{ab}^{rs}\tau_{cd}^{pq} + t_{cd}^{pq}\tau_{ab}^{rs} \right ) . \nonumber
\end{eqnarray}
%
Collecting all the diagrams, at the second order, there are 56 Goldstone 
diagrams in the $\mathbf{T}_2^{(1)}$ equation after contraction of the
cluster operators with  $H_N$.
\begin{figure}[h]
 \includegraphics[width=8cm]{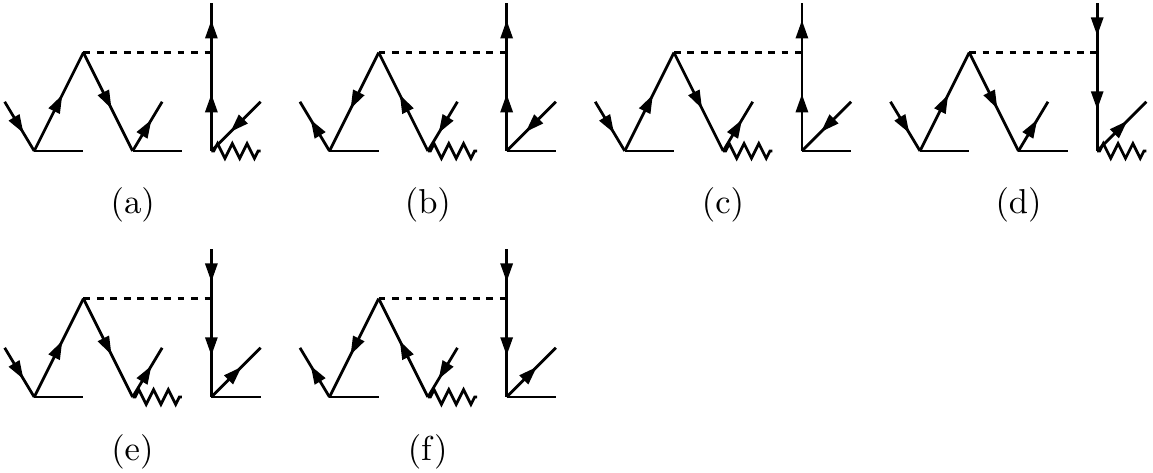}
 \caption{Diagrams arising from the contraction
 $\contraction[0.5ex]{}{H}{_{\rm N}}{T}
  \contraction[0.8ex]{}{H}{_{\rm N}T^{(0)}}{T} 
  \contraction[1.1ex]{}{H}{_{\rm N}T^{(0)}T^{(0)}}{T} 
  H_{\rm N}T_1^{(0)}T_1^{(0)}\mathbf{T}_1^{(1)} $}
 \label{dt10t10t11}
\end{figure}

\begin{figure}[h]
 \includegraphics[width=8cm]{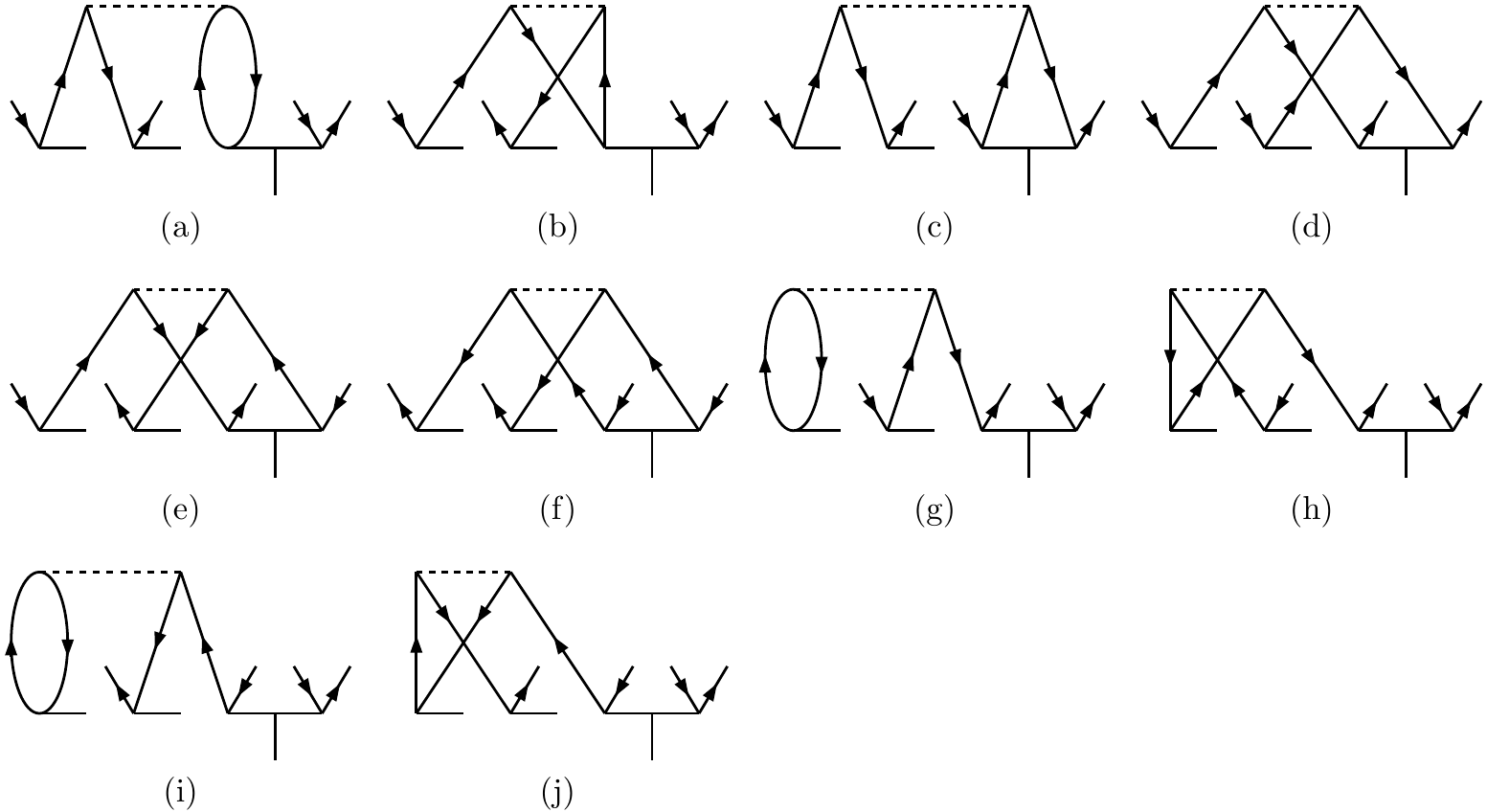}
 \caption{Diagrams arising from the contraction
 $\contraction[0.5ex]{}{H}{_{\rm N}}{T}
  \contraction[0.8ex]{}{H}{_{\rm N}T^{(0)}}{T} 
  \contraction[1.1ex]{}{H}{_{\rm N}T^{(0)}T^{(0)}}{T} 
  H_{\rm N}T_1^{(0)}T_1^{(0)}\mathbf{T}_2^{(1)} $}
 \label{dt10t10t21}
\end{figure}

\begin{figure}[h]
 \includegraphics[width=8cm]{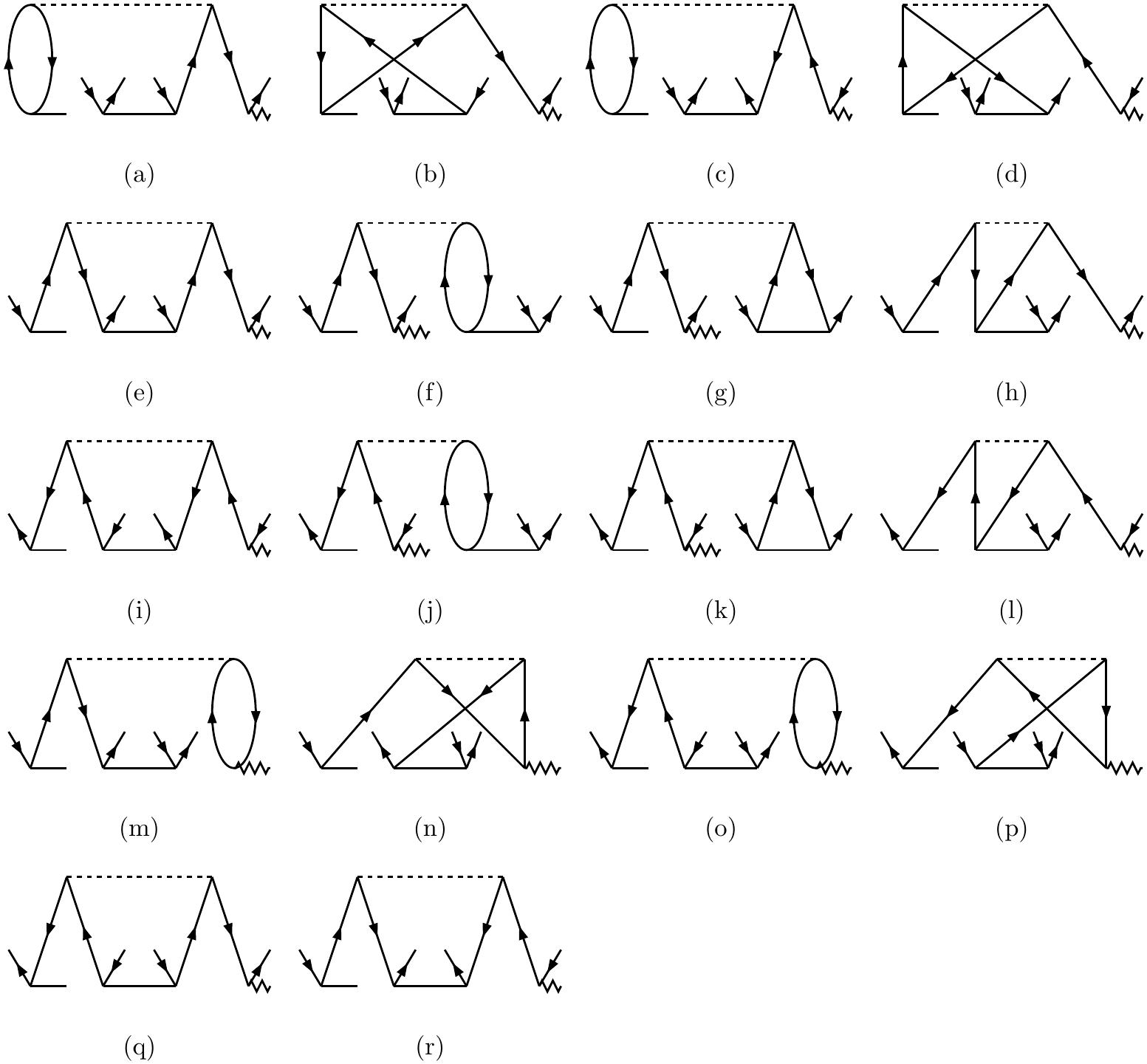}
 \caption{Diagrams arising from the contraction
  $\contraction[0.5ex]{}{H}{_{\rm N}}{T}
  \contraction[0.8ex]{}{H}{_{\rm N}T^{(0)}}{T} 
  \contraction[1.1ex]{}{H}{_{\rm N}T^{(0)}T^{(0)}}{T} 
  H_{\rm N}T_1^{(0)}T_2^{(0)}\mathbf{T}_1^{(1)} $ }
  \label{dt10t20t11}
\end{figure}

\begin{figure}[h]
 \includegraphics[width=4cm]{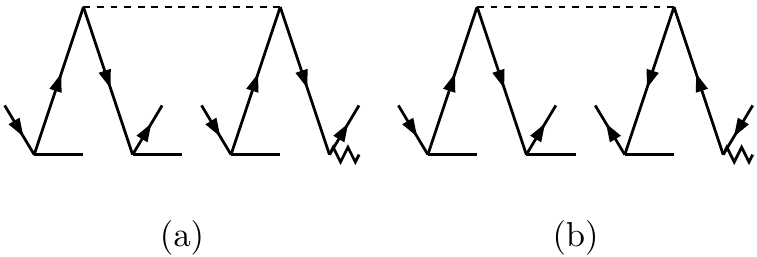}
 \caption{Diagrams arising from the contraction
 $\contraction[0.5ex]{}{H}{_{\rm N}}{T}
  \contraction[0.8ex]{}{H}{_{\rm N}T^{(0)}}{T} 
  \contraction[1.1ex]{}{H}{_{\rm N}T^{(0)}T^{(0)}}{T} 
  \contraction[1.4ex]{}{H}{_{\rm N}T^{(0)}T^{(0)}T^{(0)}}{T} 
  H_{\rm N}T_1^{(0)}T_1^{(0)}T_1^{(0)}\mathbf{T}_1^{(1)} $}
  \label{dt10t10t10t11}
\end{figure}

  At the third order, all the terms in Eq. (\ref{3rd_ord}) have non-zero
contributions to $\mathbf{T}_2^{(1)}$. There are six Goldstone diagrams from the
first term $\contraction[0.5ex]{}{H}{_{\rm N}}{T}
  \contraction[0.8ex]{}{H}{_{\rm N}T^{(0)}}{T} 
  \contraction[1.1ex]{}{H}{_{\rm N}T^{(0)}T^{(0)}}{T} 
  H_{\rm N}T_1^{(0)}T_1^{(0)}\mathbf{T}_1^{(1)} $  and these are shown in 
Fig. \ref{dt10t10t11}. The equivalent algebraic expression of the diagrams is
\begin{eqnarray}
   \langle\contraction[0.5ex]{}{H}{_{\rm N}}{T}
   \contraction[0.8ex]{}{H}{_{\rm N}T^{(0)}}{T} 
   \contraction[1.1ex]{}{H}{_{\rm N}T^{(0)}T^{(0)}}{T} 
   H_{\rm N}T_1^{(0)}T_1^{(0)}\mathbf{T}_1^{(1)}\rangle_{ab}^{pq} & = & 
   \sum_{crs} g_{cqrs} \left [- t_{a}^{r}t_{c}^{p}\tau_{b}^{s}
   - ( t_{c}^{p}\tau_{a}^{r} - t_{a}^{r}\tau_{c}^{p}) t_{b}^{s} \right ]  
                            \nonumber \\
   & & + \sum_{cdr}g_{cdrb} \left [ t_{a}^{r}(t_{c}^{p}\tau_{d}^{q}
   + \tau_{c}^{p}t_{d}^{q}) + t_{c}^{p}\tau_{a}^{r}t_{d}^{q} \right ].  \nonumber
\end{eqnarray}
The overall contribution from these diagrams is expected to be small as
these are quadratic in $T_1^{(0)}$. The next term, 
$\contraction[0.5ex]{}{H}{_{\rm N}}{T}
  \contraction[0.8ex]{}{H}{_{\rm N}T^{(0)}}{T} 
  \contraction[1.1ex]{}{H}{_{\rm N}T^{(0)}T^{(0)}}{T} 
  H_{\rm N}T_1^{(0)}T_1^{(0)}\mathbf{T}_2^{(1)} $, has ten Goldstone diagrams 
and these are shown in Fig. \ref{dt10t10t21}.
The equivalent algebraic expression of the diagrams is
\begin{eqnarray}
  \langle \contraction[0.5ex]{}{H}{_{\rm N}}{T}
  \contraction[0.8ex]{}{H}{_{\rm N}T^{(0)}}{T} 
  \contraction[1.1ex]{}{H}{_{\rm N}T^{(0)}T^{(0)}}{T} 
  H_{\rm N}T_1^{(0)}T_1^{(0)}\mathbf{T}_2^{(1)}\rangle_{ab}^{pq} & =& 
  \sum_{cdrs} g_{cdrs} \left [  t_{a}^{r}t_{c}^{p}\tilde{\tau}_{bd}^{sq}
  + t_{a}^{r}t_{d}^{p}\tau_{cb}^{sq}  + t_{a}^{r}t_{b}^{s}\tau_{cd}^{pq} \right .
                     \nonumber \\
 & & + t_{d}^{q}( t_{a}^{r}\tau_{cb}^{ps} + t_{c}^{p}\tau_{ab}^{rs} )
     - (t_{c}^{r}t_{a}^{s} - t_{c}^{s}t_{a}^{r}) \tau_{db}^{pq}
                      \nonumber \\
 & &\left.  - (t_{c}^{r}t_{d}^{p} - t_{d}^{r}t_{c}^{p})\tau_{ab}^{sq} \right ] .
                      \nonumber
\end{eqnarray}
Contributions from these diagrams will be lower than the previous set as 
these depend on $\mathbf{T}_2^{(1)} $, which is smaller in magnitude than 
$\mathbf{T}_1^{(1)}$, and quadratic in $T_1^{(0)}$ like the previous set. 
The last third order term, 
$\contraction[0.5ex]{}{H}{_{\rm N}}{T}
  \contraction[0.8ex]{}{H}{_{\rm N}T^{(0)}}{T} 
  \contraction[1.1ex]{}{H}{_{\rm N}T^{(0)}T^{(0)}}{T} 
  H_{\rm N}T_1^{(0)}T_2^{(0)}\mathbf{T}_1^{(1)} $, has eighteen diagrams and
these are shown in Fig. \ref{dt10t20t11}. 
The algebraic equivalent of these diagrams is 
\begin{eqnarray}
  \langle \contraction[0.5ex]{}{H}{_{\rm N}}{T}
  \contraction[0.8ex]{}{H}{_{\rm N}T^{(0)}}{T} 
  \contraction[1.1ex]{}{H}{_{\rm N}T^{(0)}T^{(0)}}{T} 
  H_{\rm N}T_1^{(0)}T_2^{(0)}\mathbf{T}_1^{(1)}\rangle_{ab}^{pq} & = & 
  \sum_{cdrs} g_{cdrs} \left [ ( t_{c}^{s}t_{ab}^{pr} - t_{c}^{r}t_{ab}^{ps}) 
  \tau_{d}^{q} - ( t_{c}^{r}t_{ad}^{pq}      \right .
                      \nonumber \\
  & & - t_{d}^{r}t_{ac}^{pq}) \tau_{b}^{s} + t_{a}^{r}( t_{cb}^{ps}\tau_{d}^{q} 
   - \tilde{t}_{db}^{sq}\tau_{c}^{p} + t_{cb}^{sq}\tau_{d}^{p} 
  \nonumber \\
  & & - t_{cb}^{pq}\tau_{d}^{s} + t_{db}^{pq}\tau_{c}^{s} 
   + t_{cd}^{pq}\tau_{b}^{s} ) + t_{c}^{p}( t_{ad}^{rq}\tau_{b}^{s} 
  \nonumber \\
  & & - \tilde{t}_{db}^{sq}\tau_{a}^{r} + t_{db}^{rq}\tau_{a}^{s}
   - t_{ab}^{rq}\tau_{d}^{s} + t_{ab}^{sq}\tau_{d}^{r} 
    \nonumber \\
  & & \left . + t_{ab}^{rs}\tau_{d}^{q}) \right ].  \nonumber
\end{eqnarray}
Among the third order terms in the $\mathbf{T}_2^{(1)}$ equation this will be 
the leading term as it depends on $T_2^{(0)} $ and $\mathbf{T}_1^{(1)} $, the 
dominant among the unperturbed and perturbed cluster operators, respectively.
There are two Goldstone diagrams from the fourth order term and these are 
shown in Fig. \ref{dt10t10t10t11} and the algebraic expression is 
\begin{equation}
  \langle \contraction[0.5ex]{}{H}{_{\rm N}}{T}
  \contraction[0.8ex]{}{H}{_{\rm N}T^{(0)}}{T} 
  \contraction[1.1ex]{}{H}{_{\rm N}T^{(0)}T^{(0)}}{T} 
  \contraction[1.4ex]{}{H}{_{\rm N}T^{(0)}T^{(0)}T^{(0)}}{T}
  H_{\rm N}T_1^{(0)}T_1^{(0)}T_1^{(0)}\mathbf{T}_1^{(1)}\rangle_{ab}^{pq} = 
  \sum_{cdrs} g_{cdrs}t_{a}^{r}t_{c}^{p}( t_{b}^{s}\tau_{d}^{q}
  + t_{d}^{q}\tau_{b}^{s} ).   \nonumber
\end{equation}
Among all the diagrams considered so far these two diagrams will have the 
lowest contributions as these are third order in $T_1^{(0)}$. However, for 
completeness we include these in the calculations.


\subsubsection{$\contraction[0.5ex]{}{\mathbf{D}}{}{T}\mathbf{D}T^{(0)} $ 
and $\contraction[0.5ex]{}{\mathbf{D}}{}{T}
 \contraction[0.8ex]{}{\mathbf{D}}{T^{(0)}}{T} \mathbf{D}T^{(0)}T^{(0)}$ 
diagrams}

\begin{figure}[h]
 \includegraphics[width=8cm]{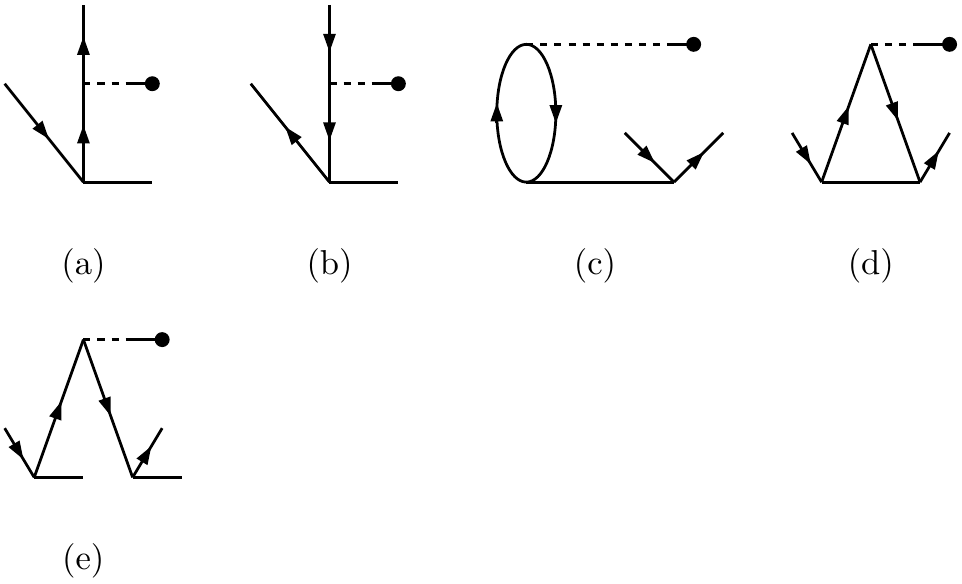}
 \caption{Singles diagrams arising from the contraction
  $\contraction[0.5ex]{}{H}{_{\rm int}}{T}H_{\rm int}T^{(0)} $ and
  $\contraction[0.5ex]{}{H}{_{\rm int}}{T}
 \contraction[0.8ex]{}{H}{_{\rm int}T^{(0)}}{T} H_{\rm int}T^{(0)}T^{(0)} $}
 \label{hint_sing} 
\end{figure}

\begin{figure}[h]
 \includegraphics[width=8cm]{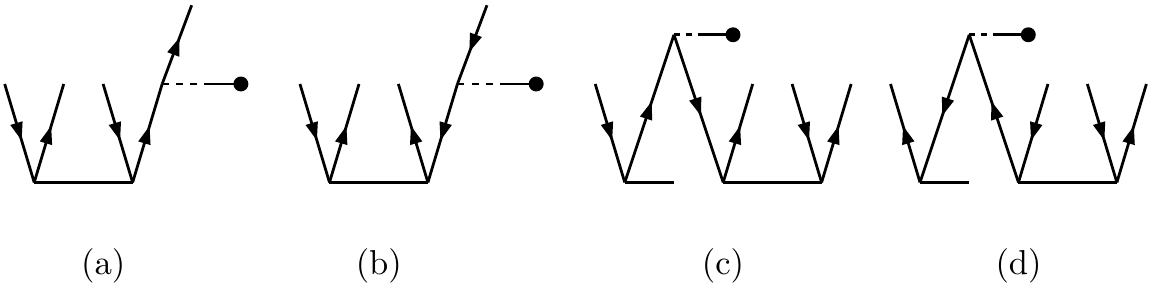}
 \caption{Doubles diagrams arising from the contraction
  $\contraction[0.5ex]{}{H}{_{\rm int}}{T}H_{\rm int}T^{(0)} $ and
  $\contraction[0.5ex]{}{H}{_{\rm int}}{T}
 \contraction[0.8ex]{}{H}{_{\rm int}T^{(0)}}{T} H_{\rm int}T^{(0)}T^{(0)} $} 
 \label{hint_dbl}
\end{figure}

 Another group of $\mathbf{T}^{(1)}$ diagrams arise from the contraction of 
$\mathbf{D}$ and $T^{(0)}$, which are present on the right hand side of 
Eq. (\ref{prcc_sing}) and (\ref{prcc_dbl}). In this group, for the 
$\mathbf{T}_1^{(1)}$ equation there are five Goldstone diagrams and these are 
shown in Fig. \ref{hint_sing}. 
However, among these only the last one is nonlinear in $T^{(0)}$. The algebraic 
expression of the diagrams is
\begin{eqnarray}
  \langle \contraction[0.5ex]{}{\mathbf{D}}{}{T}\mathbf{D}T^{(0)} \rangle_a^p + 
  \langle \contraction[0.5ex]{}{\mathbf{D}}{}{T}
  \contraction[0.8ex]{}{\mathbf{D}}{T^{(0)}}{T} \mathbf{D}T^{(0)}T^{(0)}
  \rangle_a^p & = & \sum_{q}\mathbf{r}_{pq} t_{a}^{q}
       - \sum_{c}\mathbf{r}_{ca}  t_{c}^{p}
  \nonumber \\
  &  & \sum_{bq}\mathbf{r}_{bq} \left ( t_{ba}^{qp} - t_{ab}^{qp} 
      - t_{a}^{q}t_{b}^{q} \right )  , \nonumber
\end{eqnarray}
where, $\mathbf{r}_{ij} = \langle i|\mathbf{r}|j\rangle$ is the electronic part 
of the single particle matrix element. For $\mathbf{T}_2^{(1)} $, there are four
diagrams and these are shown in Fig. \ref{hint_dbl} and last two are 
nonlinear in $T^{(0)} $ .
The algebraic expression of the diagrams is
\begin{eqnarray}
  \langle \contraction[0.5ex]{}{\mathbf{D}}{}{T}\mathbf{D}T^{(0)} 
  \rangle_{ab}^{pq} + \langle \contraction[0.5ex]{}{\mathbf{D}}{}{T}
  \contraction[0.8ex]{}{\mathbf{D}}{T^{(0)}}{T} \mathbf{D}T^{(0)}T^{(0)}
  \rangle_{ab}^{pq} & = & \sum_{r}\mathbf{r}_{qr}t_{ab}^{pr} 
       - \sum_{c}\mathbf{r}_{cb}t_{ac}^{pq}
  \nonumber \\
  &  & \sum_{cr}\mathbf{r}_{cr} \left ( - t_{a}^{r} t_{cb}^{pq} 
  - t_{c}^{p}t_{ab}^{rq} \right ) . \nonumber
\end{eqnarray}
This completes the diagrammatic and algebraic analysis of the nonlinear terms 
in the $\mathbf{T}^{(1)}$ equations. To obtain the linear algebraic equations 
of the cluster amplitudes, each of the diagrams or terms in the algebraic 
expression requires further simplification to radial and angular components. 
The angular part is evaluated diagrammatically but the diagrams are different 
from the Goldstone diagrams.


\section{Dipole Polarizability}
From the second order time-independent perturbation theory, the 
ground state dipole polarizability of a closed-shell atom is 
\begin{equation}
  \alpha = -2  \sum_{I} \frac
  {\langle \Psi_0|\mathbf D|\Psi_I\rangle \langle \Psi_I|\mathbf D|
  \psi_0\rangle}{E_0 - E_I}, 
\end{equation}
where $|\Psi_I \rangle $ are the intermediate atomic states and $E_i$ is the 
energy of the atomic state. As $\mathbf{D}$ is an odd parity operator, 
$|\Psi_I\rangle$ must be opposite in parity to $|\Psi_0\rangle$. In the PRCC 
theory we can write 
\begin{equation}
  \alpha = -\frac{\langle\tilde{\Psi}_0|\mathbf{D}|\tilde{\Psi}_0\rangle}
           {\langle\tilde{\Psi}_0|\tilde{\Psi}_0\rangle}.
\end{equation}
From the definition of $|\tilde{\Psi}_0\rangle$ in Eq. (\ref{psi_tilde}) and
based on the parity selection rules, only the terms linear in $\lambda$ 
are nonzero. That is,
\begin{equation}
  \alpha = -\frac{\langle \Phi_0|\mathbf{T}^{(1)\dagger}\bar{\mathbf{D}} + 
   \bar{\mathbf{D}}\mathbf{T}^{(1)}|\Phi_0\rangle}{\langle\Psi_0|\Psi_0\rangle},
\end{equation}
where, $\bar{\mathbf{D}} = e^{{T}^{(0)\dagger}}\mathbf{D} e^{T^{(0)}}$, 
represents the unitary transformed electric dipole operator. Retaining the 
the leading order terms, we obtain
\begin{eqnarray}
 \alpha & \approx & \frac{1}{\cal N}\langle\Phi_0|
     \mathbf{T}_1^{(1)\dagger}\mathbf{D} + \mathbf{D}\mathbf{T}_1^{(1)} 
     + \mathbf{T}_1^{(1)\dagger}\mathbf{D}T_1^{(0)} 
     + T_1^{(0)\dagger}\mathbf{D}\mathbf{T}_1^{(1)}\nonumber \\
    &&  + \mathbf{T}_2^{(1)\dagger}\mathbf{D}T_1^{(0)} 
     + T_1^{(0)\dagger}\mathbf{D}\mathbf{T}_2^{(1)}
     + \mathbf{T}_1^{(1)\dagger}\mathbf{D}T_2^{(0)}\nonumber \\ 
    && + T_2^{(0)\dagger}\mathbf{D}\mathbf{T}_1^{(1)}
     + \mathbf{T}_2^{(1)\dagger}\mathbf{D}T_2^{(0)} 
     + T_2^{(0)\dagger}\mathbf{D}\mathbf{T}_2^{(1)}
     |\Phi_0\rangle, 
 \label{exp_alpha}
\end{eqnarray}
where ${\cal N} = \langle\Phi_0|\exp[T^{(0)\dagger}]\exp[T^{(0)}]
|\Phi_0\rangle$ is the normalization factor, which involves a non-terminating
series of contractions between ${T^{(0)}}^\dagger $ and $T^{(0)} $. However, in 
the present work we use 
${\cal N} \approx \langle\Phi_0|T_1^{(0)\dagger}T_1^{(0)} + 
T_2^{(0)\dagger}T_2^{(0)}|\Phi_0\rangle$. From the above expression of $\alpha$,
an evident advantage of calculation using PRCC theory is the absence of
summation over $|\Psi_I\rangle $. The summation is subsumed in the 
evaluation of the $\mathbf{T}^{(1)}$ in a natural way. This is 
one of the key calculational advantage of using PRCC theory. 

\begin{table}[h]
        \caption{Comparison between GTO and GRASP92}
        \label{orb}
        \begin{center}
        \begin{tabular}{ldd}
            \hline
            Atom & 
              \multicolumn{1}{c}{GTO} &
              \multicolumn{1}{c}{GRASP92} \\
            \hline
            Ar & -528.6837   & -528.6837   \\
            Kr & -2789.8605  & -2788.8605  \\
            Xe & -7446.8976  & -7446.8976  \\
            Rn & -23602.0202 & -23602.0232 \\
           \hline
        \end{tabular}
        \end{center}
\end{table}


\section{Calculational details}

\subsection{Basis set and nuclear density}
 The first step of our calculations, which is also true of any atomic and 
molecular calculations, is to generate an orbital basis set. For the present
work, we use the Dirac-Hartree-Fock Hamiltonian and even-tempered Gaussian type 
orbitals (GTOs) \cite{mohanty-90}. The radial part of the spin-orbitals, the 
large component in particular, are linear combinations of the Gaussian type 
functions
\begin{equation}
   g_{\kappa p}^{L}(r) = C^{L}_{\kappa i} r^{n_{\kappa}}e^{-\alpha_{p}r^{2}},
\end{equation}
where $p$ is the GTO index and $C^{L}_{\kappa i}$ is the normalization 
constant. The exponent $\alpha_{p}$ depends on two parameters 
$\alpha_{0}$ and $\beta$, these are related as 
$\alpha_{p} = \alpha_{0} \beta^{p-1}$, where $p=0,1\ldots m$ and $m$ is the 
number of the gaussian type functions. The small components of the 
spin-orbitals are linear combination of $g_{\kappa p}^{S}(r)$, which are 
generated from $g_{\kappa p}^{L}(r)$ through the kinetic balance condition 
\cite{stanton-84}. We calculate the GTOs on a grid \cite{chaudhuri-99} and 
optimize the values of $\alpha_{0}$ and $\beta$ for individual atoms to match 
the spin-orbital energies and self consistent field (SCF) energy of GRASP92 
\cite{parpia-96}, which solves Dirac-Hartree-Fock equations numerically. 
The comparison of the SCF energies are given in Table. \ref{orb}. Except 
for Rn, there is excellent agreement between the SCF energies obtained from 
GTO and GRASP92. The symmetry wise values of the optimized $\alpha_{0}$ 
and $\beta$ are listed in Table. \ref{basis}.
\begin{table}[h]
   \caption{The $\alpha_0$ and $\beta$ parameters of the even tempered
            GTO basis used in the present calculations.}
   \label{basis}
   \begin{tabular}{cccccccc}
   \hline
   \hline
     Atom & \multicolumn{2}{c}{$s$} & \multicolumn{2}{c}{$p$} & 
     \multicolumn{2}{c}{$d$}  \\
     & $\alpha_{0}$  & $\beta$ & $\alpha_{0}$ & $\beta$  
     & $\alpha_{0}$  & $\beta$  \\
     \hline
     Ar &\, 0.00055  &\, 1.620 &\, 0.00515 &\, 2.405 &\, 0.00570 &\, 2.850 \\
     Kr &\, 0.00015  &\, 2.015 &\, 0.00945 &\, 2.975 &\, 0.00635 &\, 2.845 \\
     Xe &\, 0.00012  &\, 2.215 &\, 0.00495 &\, 2.995 &\, 0.00745 &\, 2.460 \\
     Rn &\, 0.00010  &\, 2.280 &\, 0.00671 &\, 2.980 &\, 0.00715 &\, 2.720 \\
     \hline
   \end{tabular}
\end{table}
To optimize the basis set size, we examine the convergence of $\alpha $ using
the LPRCC theory. We start with a basis set of 50 GTOs and increase 
the basis set size in steps through a series of calculations. As an example
the results for the case of Kr is listed in Table. \ref{basis_kr}. The value
of $\alpha$ changes by only $7\times 10^{-4}$ when the number of basis states
is increased from 117 to 131. So, we can use the former for our calculations
without compromising the desired accuracy.
\begin{table}[h]
  \caption{Convergence pattern of $\alpha$ (Kr) as a function of
           the basis set size.}
  \label{basis_kr}
  \begin{tabular}{lcc}
      \hline
      No. of orbitals & Basis size & $\alpha $   \\
      \hline
      79  & $(15s, 9p,  9d,   7f,   7g)  $ & 16.8759  \\
      97  & $(17s, 11p, 11d,  9f,   9g)  $ & 16.7507  \\
      117 & $(21s, 13p, 13d,  11f,  11g) $ & 16.7403  \\
      131 & $(25s, 15p, 14d,  13f,  11g) $ & 16.7396  \\
      139 & $(25s, 16p, 15d,  13f,  13g) $ & 16.7394  \\
      155 & $(29s, 17p, 16d,  15f,  15g) $ & 16.7394  \\
      \hline
  \end{tabular}
\end{table}

  In the present work we have considered finite size Fermi density 
distribution of the nucleus
\begin{equation}
   \rho_{\rm nuc}(r) = \frac{\rho_0}{1 + e^{(r-c)/a}},
\end{equation}
where, $a=t 4\ln(3)$. The parameter $c$ is the half charge radius so that 
$\rho_{\rm nuc}(c) = {\rho_0}/{2}$ and $t$ is the skin thickness. The PRCC 
equations are solved iteratively using Jacobi method, we have chosen this 
method as it is easily parallelizable. The method, however, is slow to 
converge. So, we use direct inversion in the iterated subspace (DIIS)
\cite{pulay-80} to accelerate the convergence.


\subsection{Breit interaction}
There are two different but equivalent approaches, reported in previous 
works, to calculate the matrix elements of $g^{\rm B}(r_{12})$. The first 
approach \cite{mann-71} is to couple the angular part of the orbitals with 
$\bm{\alpha}$ to give a linear combination of vector spherical harmonics. This 
is then combined with the 
angular part of $1/r_{12}$ for integration. In the second approach 
\cite{grant-76}, $g^{\rm B}(r_{12})$ is expanded as a linear combination of 
irreducible tensor operators. In the present work we use the later and 
employ the expressions given in ref. \cite{grant-80} to incorporate 
$g^{\rm B}(r_{12}) $ in the GTO and RCC calculations. For GTO calculation
\cite{quiney-03,clementi-91} provides a very good description to include
$g^{\rm B}(r_{12}$ in finite basis set calculations. To assess the relative 
importance of Breit interaction, we calculate the first order energy correction 
\begin{equation}
  \langle H^{\rm B}\rangle_{\rm DF} = \langle\Phi_0|\sum_{i<j}g^{\rm B}(r_{ij})
       |\Phi_0\rangle,
\end{equation}
where, $|\Phi_o\rangle $ is the ground state reference function generated from 
the Dirac-Hartree-Fock spin-orbitals and 
$H^{\rm B} =\sum_{i<j}g^{\rm B}(r_{ij})$ represents the many-particle form of 
the Breit interaction. The $\langle H^{\rm B}\rangle_{\rm DF}$ of the 
rare gas atoms Ar, Kr, Xe and Rn are listed in Table. \ref{scf_energy}.
\begin{table}[h]
  \caption{SCF Energies for noble gas atoms}
  \label{scf_energy}
  \begin{tabular}{ldddd}
  \hline
    Atom    & \multicolumn{1}{c}{$E_{\rm SCF}^{\rm DC}$}  &
              \multicolumn{1}{c}{$E_{\rm SCF}^{\rm DCB}$} &
              \multicolumn{1}{c}{$\langle H^{\rm B}\rangle_{\rm DF} $} &
              \multicolumn{1}{c}{Ref. \cite{parpia-92}}   \\ \hline
    Ar   & -528.6837   &  -528.5511    & 0.1326    &  0.1324       \\
    Kr   & -2788.8605  &  -2787.4310   & 1.4295    &  1.4268       \\
    Xe   & -7446.8976  &  -7441.1248   & 5.7728    &  5.7753       \\
    Rn   & -23602.0202 &  -23572.8480  & 29.1722   &  29.3968      \\
  \hline
  \end{tabular}
\end{table}
For each atom we calculated the SCF energy with $H^{\rm DC}$ and
$H^{\rm DCB}$, these are  
$E_{\rm SCF}^{\rm DC}= \langle\Phi_0|H^{\rm DC}|\Phi_0\rangle$ and
$E_{\rm SCF}^{\rm DCB}= \langle\Phi_0|H^{\rm DCB}|\Phi_0\rangle$. Here,
$H^{\rm DC}= H^{\rm DCB}-H^{\rm B}$, the atomic Hamiltonian without the Breit 
interaction. From the table, it is evident that our results
are in very good agreement with the previous results  \cite{parpia-92}. The
largest deviation is observed in Rn, our result of 
$\langle H^{\rm B}\rangle_{\rm DF} $ 
is 0.8\% lower than the previous result. However, as the 
Breit interaction contribution to $E_{\rm SCF}^{\rm DCB}$ is a mere 0.12\% in 
Rn, in absolute terms, the deviation is $\approx 0.001$\%. Our results are
also in good agreement with the results of another previous study 
\cite{ishikawa-91}. In the PRCC calculations, as described earlier, we treat 
$ H^{\rm B}$ at par with the residual Coulomb interaction. However, to examine 
the relative importance of Breit interaction, we calculate $\alpha$ with and without $H^{\rm B}$.


\section{Results and Discussions}

 The expression of $\alpha$ in PRCC theory, as mentioned earlier, is a 
non-terminating series. However, the terms of order higher than quadratic 
in $T$ have negligible contributions. For this reason, in the present work, we 
consider upto second order in cluster operator. As mentioned earlier, we have
added Breit interaction in the total atomic Hamiltonian, one immediate outcome 
is, the number of two-electron integrals is larger and storing these, for faster
computations, require larger memory. At the first order MBPT, which we use as 
the initial guess, there is an important change with the inclusion of 
$H^{\rm B} $. With only the Coulomb interaction, at the first order MBPT, 
the wave operator follows the parity selection rule and only selected 
multipoles of the Coulomb interaction contributes. However, with 
$H^{\rm B} $, which has opposite parity selection rule compared to Coulomb 
interaction, all multipoles of the two-electron interaction which satisfy 
the triangular conditions are allowed. In table \ref{l_pol}, we 
list the values of $\alpha$ calculated using the LPRCC theory. For 
comparison we have also included the results from previous theoretical 
studies and experimental data. There are no discernible trends in the 
previous theoretical results and present work. For Kr and Xe, the results
from the many-body perturbation theory (MBPT) \cite{thakkar-92} is higher
than the experimental data, but with RCCSD triples (RCCSDT) approximations
\cite{nakajima-01}, Ar  and Kr have higher values. For Ar our result is 1\%
higher than the experimental data and this is consistent with the 
RCCSDT result reported in a previous work. It must, however, be mentioned that 
the previous work is based on third-order Douglas-Kroll \cite{nakajima-00}
method. Our result for Kr is in excellent agreement with the experimental 
data. This could be a coincidence arising from  well chosen basis set 
parameters and inherent property of PRCC to incorporate correlation effects
more completely within a basis set. 
\begin{table}[h]
  \caption{The static dipole polarizability, $\alpha$ (atomic units), from
           linearized PRCC  and comparison with previous results}
  \label{l_pol}
  \begin{tabular}{lddddd}
  \hline
    Method  & \multicolumn{1}{c}{Ar}  &  \multicolumn{1}{c}{Kr}  &
              \multicolumn{1}{c}{Xe}  &  \multicolumn{1}{c}{Rn}  \\
    \hline \hline
    RCCSDT\cite{nakajima-01} &  11.22     &  16.80     & 27.06   & 33.18    \\
    CCSDT \cite{soldan-01}   &  11.084    &  16.839    & 27.293  & 34.43    \\
    MBPT\cite{thakkar-92}    &  11.062    &  17.214    & 28.223  &          \\
    This work                &  11.213    &  16.736    & 26.432  & 35.391   \\
    Expt.\cite{langhoff-69}  &  11.091    &  16.740    & 27.340  &          \\
    Expt.\cite{orcutt-67}    &  11.081(5) &  16.766(8) &         &          \\
  \hline
  \end{tabular}
\end{table}
 In the case of Xe our result is 3.4\% lower than the experimental data and 
2.4 \% lower than the RCCSDT result. The later, difference from the the RCCSDT 
result, can be partly attributed to the triple excitations. There is no 
experimental data of $\alpha$ for Rn, the 
highest $Z$ atom among the noble gases. In ref. \cite{nakajima-01}, 
the $\alpha $ of Rn is computed using RCCSDT and their result is 6.2\% lower
than our result. 

To estimate the importance of Breit interaction, we exclude $H^{\rm B} $ in 
the PRCC calculations and the value after excluding $H^{\rm B}$ are 11.202,
16.728, 26.404, 35.266 for Ar, Kr, Xe and Rn respectively. These represent a 
decrease of 0.010, 0.012, 0.021 and 0.133 from the results with the exclusion
of $H^{\rm B} $. Except for Rn, the change in $\alpha$ is below $0.1$\%. 
This implies that to obtain accurate results for Rn, it is desirable to 
include Breit interaction in the calculations.
\begin{table}[h]
    \caption{Contribution to $\alpha $ from different terms of the
             dressed dipole operator in the linearized PRCC theory}
    \label{result_lprcc}
    \begin{center}
    \begin{tabular}{ldddd}
        \hline
        Contributions from  & \multicolumn{1}{c}{Ar} & \multicolumn{1}{c}{Kr} 
        & \multicolumn{1}{c}{Xe} & \multicolumn{1}{c}{Rn}  \\
        \hline
        $\mathbf{T}_1^{(1)\dagger}\mathbf{D} $ + h.c.
        & 12.191 & 18.613 & 30.855 & 41.641 \\
        $\mathbf{T}_1{^{(1)\dagger}} \mathbf{D}T_2^{(0)} $  + h.c. 
        & -0.545 & -0.888 & -1.677 & -2.328 \\
        $\mathbf{T}_2{^{(1)\dagger}}\mathbf{D}T_2^{(0)} $ + h.c.  
        & 0.510 & 0.748 & 1.352 & 1.862 \\
        $\mathbf{T}_1{^{(1)\dagger}}\mathbf{D}T_1^{(0)} $ + h.c.  
        & -0.057 & -0.118 & -0.357 & -0.301 \\
        $\mathbf{T}_2{^{(1)\dagger}}\mathbf{D}T_1^{(0)} $  + h.c.  
        & 0.022  & 0.038 & 0.092  & 0.073 \\
        Normalization & 1.081 & 1.099 & 1.145 & 1.157 \\
        Total & 11.213 & 16.736  & 26.432 & 35.391 \\
        \hline
    \end{tabular}
    \end{center}
\end{table}

To examine the results in more detail, the contributions from the
terms in the expression of $\alpha$ given in Eq. (\ref{exp_alpha}) are listed 
in Table \ref{result_lprcc}. It is evident that
$\mathbf{T}_1^{(1)\dagger}\mathbf{D}$ and it's hermitian conjugate
have the leading order contributions. This is to be expected as these terms 
include the Dirac-Hartree-Fock-Breit  contribution and RPA effects,
which have the dominant contributions. In all the cases, the result from 
$\mathbf{T}_1^{(1)\dagger}\mathbf{D}$ is larger than the total value and
shows dependence on $Z$: the results of Ar, Kr, Xe and Rn from this term are
8.7\%, 11.2\%, 16.7\% and 17.7\% higher than the total values, respectively. 
The next to leading order terms are 
$\mathbf{T}_1{^{(1)\dagger}}\mathbf{D}T_2^{(0)} $ and its hermitian 
conjugate. Contributions from these terms are, approximately, a factor of
twenty smaller than the leading order terms and opposite in phase. On 
closer inspection, it is natural that 
$\mathbf{T}_1{^{(1)\dagger}}\mathbf{D}T_2^{(0)} $ and it's hermitian 
conjugate are the next to leading order terms. Among the second order
terms, these are the ones which have $\mathbf{T}_1{^{(1)}} $  and
$ T_2^{(0)} $, the dominant cluster amplitudes in the perturbed and
unperturbed relativistic coupled-cluster theories. The results from 
$\mathbf{T}_1{^{(1)\dagger}}\mathbf{D}T_2^{(0)} $ have large cancellations
with the term $\mathbf{T}_2{^{(1)\dagger}}\mathbf{D}T_2^{(0)}$, which is 
almost the same in magnitude but opposite in sign. Interestingly, a similar 
pattern is also observed with the 
$\mathbf{T}{^{(1)\dagger}}\mathbf{D}T_2^{(0)}$  terms.  Namely, the results 
from $\mathbf{T}_1{^{(1)\dagger}}\mathbf{D}T_2^{(0)}$  are negative and 
opposite in sign to $\mathbf{T}_{2}{^{(1)\dagger}}\mathbf{D}T_2^{(0)}$.

\begin{table}[h]
    \caption{Contribution to $\alpha $ from different terms of the
             dressed dipole operator in the non-linear PRCC theory}
    \label{result_nprcc}
    \begin{center}
    \begin{tabular}{lddd}
        \hline
        Contributions from  & \multicolumn{1}{c}{Ar} & \multicolumn{1}{c}{Kr} 
        & \multicolumn{1}{c}{Xe}   \\
        \hline
        $\mathbf{T}_1^{(1)\dagger}\mathbf{D} $ + h.c.
        & 12.950 & 18.622 & 33.108  \\
        $\mathbf{T}_1{^{(1)\dagger}}\mathbf{D}T_2^{(0)} $  + h.c. 
        & -0.579 & -0.899 & -1.7964  \\
        $ \mathbf{T}_2{^{(1)\dagger}}\mathbf{D}T_2^{(0)} \}$ + h.c.
        & 0.488 & 0.769 & 1.278  \\
        $\mathbf{T}_1{^{(1)\dagger}}\mathbf{D}T_1^{(0)} $ + h.c.  
        & -0.061 & -0.096 & -0.392  \\
        $\mathbf{T}_2{^{(1)\dagger}}\mathbf{D}T_1^{(0)} $  + h.c.
        & 0.022  & 0.035 & 0.095   \\
        Normalization & 1.081 & 1.099 & 1.145  \\
        Total & 11.859 & 16.771 & 28.203  \\
        \hline
    \end{tabular}
    \end{center}
\end{table}

 The results from the full PRCC, including the terms nonlinear in cluster
amplitudes are given in table \ref{result_nprcc}. From the table, it is 
clear that the nonlinear terms tend to increase the deviations from the 
experimental data. A similar trend was reported in our previous work on
Ne \cite{chattopadhyay-12}. For Ar, the non-linear PRCC theory result is 
5.4\% larger than the result from linearized PRCC and it is 6.5\% larger
than the experimental result. Similarly, for Xe the nonlinear PRCC result
is 6.3\% larger than the linearized PRCC result. On the other hand for 
Kr, the non-linear PRCC results are marginally larger than the 
linearized PRCC results. The larger values of $\alpha$ in the non-linear 
PRCC can almost entirely be attributed to higher value of 
$ \mathbf{T}_1^{(1)\dagger}\mathbf{D} $ and it's hermitian conjugate. It
means that the non-linear terms tend to increase the RPA effects. This is 
an example where inclusion of higher order terms enhance the uncertainty. 
It is possible that triple excitations, higher order excitation
not considered in the present work, may balance the deviations and bring the 
results closer to the experimental data. 
\begin{table}[h]
    \caption{Core orbital contribution from 
        $ \mathbf{T}_1^{(1)\dagger}\mathbf{D}$ to $\alpha $}
    \label{result_t1d}
    \begin{center}
    \begin{tabular}{crrr}
        \hline
          \multicolumn{1}{c}{Ar} & \multicolumn{1}{c}{Kr}
        & \multicolumn{1}{c}{Xe} & \multicolumn{1}{c}{Rn}  \\
        \hline
        8.152 (3$p_{3/2}$) & 12.872 (4$p_{3/2}$) & 22.292 (5$p_{3/2}$) 
        & 34.524 (6$p_{3/2}$) \\
        3.914 (3$p_{1/2}$) & 5.572 (4$p_{1/2}$) & 8.120  (5$p_{1/2}$) 
        & 6.502 (6$p_{1/2}$)  \\
        0.100 (3$s_{1/2}$) & 0.058 (4$s_{1/2}$) & 0.222 (4$d_{5/2}$) 
        & 0.382 (5$d_{5/2}$)  \\
        0.012 (2$p_{3/2}$) & 0.056 (3$d_{5/2}$) & 0.140 (4$d_{3/2}$) 
        & 0.214 (5$d_{3/2}$)  \\
        \hline
    \end{tabular}
    \end{center}
\end{table}

  For a more detailed analysis of the contributions from the RPA effects, 
we consider contributions from each of the core orbitals in 
$ \mathbf{T}_1^{(1)\dagger}\mathbf{D} $. In terms of orbital indices
the expression is 
\begin{equation}
  \mathbf{T}_1^{(1)\dagger}\mathbf{D} + \text{H.c.} = 
   \sum_{a p} \left ( \mathbf{d}_{ap}\tau_a^p 
   + {\tau_a^p}^*\mathbf{d}_{pa} \right ),
\end{equation}
where, $\mathbf{d}_{ap} = \langle a|\mathbf{d}|p\rangle$ with $\mathbf{d}$ as 
the single particle electric dipole operator. The four leading core orbitals 
($a$) for each of the atoms are listed in 
Table. \ref{result_t1d}. In all the cases, the result from the outermost 
$np_{3/2}$ valence orbitals are the largest. This is not surprising as these 
are the orbitals which are spatially most extended. In addition, as the matrix 
elements in the expression of $\alpha$ has a quadratic dependence on radial 
distance, orbitals with larger radial extent have higher contributions. 
The next largest values arise from the $np_{1/2}$ valence orbitals. An 
interesting pattern is to be noticed in the results, with higher $Z$ the 
ratio of the contribution from $np_{3/2}$ to $np_{1/2}$ increases. 
For Ar, Kr and Xe the ratios are 2.1, 2.3 and 2.7, respectively. However, 
the ratio for Rn is much larger, it is 5.3. The reason for the trend in the 
ratios is the contraction of the $np_{1/2}$ core orbitals due to relativistic 
corrections. Hence, the $np_{1/2}$ valence orbitals of higher $Z$ atoms show 
larger contraction and accounts for the higher ratio. The third largest 
contributions in Ar and Kr arise from the $3s_{1/2}$ and $4s_{1/2}$ orbitals, 
respectively. This is expected as these are the orbitals which are 
energetically just below the $np$ orbitals and spatially as well. Extending the same pattern, for Xe and Rn, the third largest contributions must be from 
the $5s_{1/2}$ and $6s_{1/2}$ orbitals, respectively, but this is not case. 
These orbitals are contracted because of relativistic corrections and the 
diffused $nd_{5/2}$ orbitals have the third largest values. From the trends in 
the results of the RPA effects, it is obvious that the relativistic corrections 
are important for Xe and Rn. 
\begin{table}[h]
  \caption{Core orbitals contribution from
           $\mathbf{T}_1{^{(1)\dagger}} \mathbf{D}T_2^{(0)}$ to $\alpha$ 
           of Argon and Krypton}
    \label{t1dt2_ar_kr}
    \begin{center}
    \begin{tabular}{dcdc}
       \hline
       \multicolumn{2}{c}{Ar}  & \multicolumn{2}{c}{Kr}  \\
       \hline
       -0.124 & $(3p_{3/2}, 3p_{1/2})$  &  -0.205 & $(4p_{3/2}, 4p_{1/2})$  \\
       -0.118 & $(3p_{3/2}, 3p_{3/2})$  &  -0.193 & $(4p_{3/2}, 4p_{3/2})$  \\
       -0.027 & $(3p_{1/2}, 3p_{1/2})$  &  -0.038 & $(4p_{1/2}, 4p_{1/2})$  \\
       -0.006 & $(3p_{3/2}, 3s_{1/2})$  &  -0.008 & $(4p_{3/2}, 3d_{5/2})$  \\
     \hline
    \end{tabular}
    \end{center}
\end{table}

  Next, we examine the pair-correlation effects, which manifest through the 
next to leading order terms, $\mathbf{T}_1{^{(1)\dagger}}\mathbf{D}T_2^{(0)}$ 
and it's hermitian conjugate. In terms of orbital indices
\begin{eqnarray}
  \mathbf{T}_1{^{(1)\dagger}}\mathbf{D}T_2^{(0)} + \text{H.c.} & = &
  \sum_{abpq} \left [ \left ( {\tau_a^p}^*\mathbf{d}_{bq} 
  - {\tau_a^q}^*\mathbf{d}_{bp} \right ) t_{ab}^{pq} \right . 
                    \nonumber  \\ 
  & & + \left . {t_{ab}^{pq}}^*\left (\tau_a^p\mathbf{d}_{qb} 
  - \tau_a^q\mathbf{d}_{pb} \right ) \right ].
\end{eqnarray}
The results of the four leading terms, listed in terms of the pairs of the core 
orbitals ($ab$), for Ar and Kr are given in Table. \ref{t1dt2_ar_kr}. From
the table we can identify $(np_{3/2}, np_{1/2})$ as the most dominant pairing 
of the core-orbitals among the double excitations. Considering that the pairing
is between different orbitals, the number of cluster amplitudes is large 
and this explains the large contribution. The second and third
dominant contributions, from the $(np_{3/2}, np_{3/2}) $ and 
$(np_{1/2}, np_{1/2}) $ pairs, are also on account of number of cluster
amplitudes. Since $np_{3/2}$ and $np_{1/2}$ each accommodate four and two
electrons each, respectively, the former has a larger number of cluster 
amplitudes. There is a small but important change in the results of Xe and Rn 
listed in Table. \ref{t1dt2_xe_rn}. The most dominant pair for these atoms is 
$(np_{3/2}, np_{3/2})$ and $(np_{3/2}, np_{1/2})$ is the second. This is in 
contrast to the sequence observed in Ar and Kr. The reason
is, although the later pair has more cluster amplitudes, the $np_{1/2}$ 
is contracted due to relativistic corrections. So, the contributions to 
$\alpha$ from $T^{(0)}_2$ involving $np_{1/2} $ is smaller. The difference
between the results from the two pairs is even more dramatic in Rn. There
are other changes in the case of Rn. The $ (6p_{3/2}, 5d_{5/2})$ pair, 
involving the diffused $5d_{5/2}$, is now the third largest contribution.
And the $(6p_{1/2}, 6p_{1/2})$, which has the contracted $6p_{1/2}$ orbital,
is the fourth largest contribution. This difference in the sequence of leading 
contributions for Rn arises from the larger relativistic corrections.
\begin{table}[h]
  \caption{Core orbitals contribution from
           $\mathbf{T}_1{^{(1)\dagger}} \mathbf{D}T_2^{(0)} \}$ to 
           $\alpha$ of Xenon and Radon}
    \label{t1dt2_xe_rn}
    \begin{center}
    \begin{tabular}{dcdc}
       \hline
       \multicolumn{2}{c}{Xe}  & \multicolumn{2}{c}{Rn}  \\
       \hline
       -0.361 & $(5p_{3/2}, 5p_{3/2})$  &  -0.591 & $(6p_{3/2}, 6p_{3/2})$  \\
       -0.359 & $(5p_{3/2}, 5p_{1/2})$  &  -0.387 & $(6p_{3/2}, 6p_{1/2})$  \\
       -0.054 & $(5p_{1/2}, 5p_{1/2})$  &  -0.071 & $(6p_{3/2}, 5d_{5/2})$  \\
       -0.035 & $(5p_{3/2}, 4d_{5/2})$  &  -0.036 & $(6p_{1/2}, 6p_{1/2})$  \\
     \hline
    \end{tabular}
    \end{center}
\end{table}

 To estimate the uncertainty in our calculations, we have identified few 
important sources of uncertainty. The first one is the truncation of orbital 
basis sets. Although we start with 9 symmetry for all the calculations, we
increase the number of symmetries upto 13 in steps. The basis set chosen for
the results given are after the value of $\alpha$ converges to $10^{-4}$. So,
the uncertainty from the basis set truncation is negligible. The second source 
of uncertainty is the truncation of CC theory at the single and double
excitation for both the unperturbed and perturbed RCC theory. Based on  
earlier studies, the contributions from the triples and quadruple excitations
could be at the most $\approx$3.3\%. This is also consistent with the 
deviations from the experimental data. Finally, the truncation of 
$e^{{\mathbf{T}^{(1)}}^\dagger}\mathbf{D}e^{T^{(0)}} 
+ e^{{T^{(0)}}^\dagger}\mathbf{D}e^{\mathbf{T}^{(1)}}$ is another
source of uncertainty. From our earlier studies with iterative method 
\cite{mani-10}, to incorporate higher order terms in the properties 
calculations with CC theory, the contributions from the third or higher order 
in is negligibly small. Quantum electrodynamical (QED) corrections in this
set of calculations is another source of uncertainty. However, it is 
expected to be smaller then the correction from the Breit interaction. As the
largest Breit correction, in the case of Rn, is 0.1\%, we can assume the
corrections from QED effects to be at the most 0.1\%. So, adding this the 
maximum uncertainty in our calculations is 3.4\%. However, it must be emphasized
that, for Ar and Kr, the uncertainty is much smaller than this bound.


\section{Conclusion}

  The PRCC theory is a general extension of the RCC method to incorporate
an additional perturbation. The present work demonstrates that it is suitable 
for properties calculations for closed-shell atoms. Although, in the present
work we have used PRCC to calculate electric dipole polarizability, the method
can be extended to calculate other atomic properties as well. 

  From the present study, through the detailed analysis and identification
of the dominant contributions, the relativistic correction to $\alpha$
of noble gas atoms arising from the contraction of the outermost 
$p_{1/2}$ is significant. The notable impact of this is the higher fractional 
contribution from $np_{3/2}$ in the terms which subsume RPA effects, 
$ \mathbf{T}_1^{(1)\dagger}\mathbf{D}$ and it's hermitian conjugate, as we go 
from Ar to Rn. For Rn, the effect of relativistic corrections is also 
identifiable without ambiguity in the pair-correlation effects,
the $(6p_{1/2}, 6p_{1/2})$ pair is below the $(6p_{3/2}, 5d_{5/2})$ pair
for $\mathbf{T}_1{^{(1)\dagger}}\mathbf{D}T_2^{(0)} $.

  We have also examined the importance of Breit interaction in the 
calculation of $\alpha$. The largest change of 0.1\% is associated with  Rn,
the heaviest noble gas atom. So, when the required uncertainty of the 
calculations is below 1\%, the inclusion of Breit interaction is 
desirable for higher $Z$ closed-shell atoms like Rn.


\begin{acknowledgements}
We thank S. Gautam, Arko Roy and Kuldeep Suthar for useful discussions. The
results presented in the paper are based on the computations using the 3TFLOP
HPC Cluster at Physical Research Laboratory, Ahmedabad.
\end{acknowledgements}

\bibliography{references}{}

\begin{thebibliography}{42}%
\makeatletter
\providecommand \@ifxundefined [1]{%
 \@ifx{#1\undefined}
}%
\providecommand \@ifnum [1]{%
 \ifnum #1\expandafter \@firstoftwo
 \else \expandafter \@secondoftwo
 \fi
}%
\providecommand \@ifx [1]{%
 \ifx #1\expandafter \@firstoftwo
 \else \expandafter \@secondoftwo
 \fi
}%
\providecommand \natexlab [1]{#1}%
\providecommand \enquote  [1]{``#1''}%
\providecommand \bibnamefont  [1]{#1}%
\providecommand \bibfnamefont [1]{#1}%
\providecommand \citenamefont [1]{#1}%
\providecommand \href@noop [0]{\@secondoftwo}%
\providecommand \href [0]{\begingroup \@sanitize@url \@href}%
\providecommand \@href[1]{\@@startlink{#1}\@@href}%
\providecommand \@@href[1]{\endgroup#1\@@endlink}%
\providecommand \@sanitize@url [0]{\catcode `\\12\catcode `\$12\catcode
  `\&12\catcode `\#12\catcode `\^12\catcode `\_12\catcode `\%12\relax}%
\providecommand \@@startlink[1]{}%
\providecommand \@@endlink[0]{}%
\providecommand \url  [0]{\begingroup\@sanitize@url \@url }%
\providecommand \@url [1]{\endgroup\@href {#1}{\urlprefix }}%
\providecommand \urlprefix  [0]{URL }%
\providecommand \Eprint [0]{\href }%
\providecommand \doibase [0]{http://dx.doi.org/}%
\providecommand \selectlanguage [0]{\@gobble}%
\providecommand \bibinfo  [0]{\@secondoftwo}%
\providecommand \bibfield  [0]{\@secondoftwo}%
\providecommand \translation [1]{[#1]}%
\providecommand \BibitemOpen [0]{}%
\providecommand \bibitemStop [0]{}%
\providecommand \bibitemNoStop [0]{.\EOS\space}%
\providecommand \EOS [0]{\spacefactor3000\relax}%
\providecommand \BibitemShut  [1]{\csname bibitem#1\endcsname}%
\let\auto@bib@innerbib\@empty
\bibitem [{\citenamefont {Khriplovich}(1991)}]{khriplovich-91}%
  \BibitemOpen
  \bibfield  {author} {\bibinfo {author} {\bibfnamefont {I.}~\bibnamefont
  {Khriplovich}},\ }\href {http://books.google.co.in/books?id=QjfeiNBrFC4C}
  {\emph {\bibinfo {title} {Parity Nonconservation in Atomic Phenomena}}}\
  (\bibinfo  {publisher} {Gordon and Breach Science Publishers, Philadelphia},\
  \bibinfo {year} {1991})\BibitemShut {NoStop}%
\bibitem [{\citenamefont {Udem}\ \emph {et~al.}(2002)\citenamefont {Udem},
  \citenamefont {Holzwarth},\ and\ \citenamefont {Hansch}}]{udem-02}%
  \BibitemOpen
  \bibfield  {author} {\bibinfo {author} {\bibfnamefont {T.}~\bibnamefont
  {Udem}}, \bibinfo {author} {\bibfnamefont {R.}~\bibnamefont {Holzwarth}}, \
  and\ \bibinfo {author} {\bibfnamefont {T.~W.}\ \bibnamefont {Hansch}},\
  }\href {\doibase http://dx.doi.org/10.1038/416233a} {\bibfield  {journal}
  {\bibinfo  {journal} {Nature}\ }\textbf {\bibinfo {volume} {416}},\ \bibinfo
  {pages} {233} (\bibinfo {year} {2002})}\BibitemShut {NoStop}%
\bibitem [{\citenamefont {Diddams}\ \emph {et~al.}(2004)\citenamefont
  {Diddams}, \citenamefont {Bergquist}, \citenamefont {Jefferts},\ and\
  \citenamefont {Oates}}]{diddams-04}%
  \BibitemOpen
  \bibfield  {author} {\bibinfo {author} {\bibfnamefont {S.~A.}\ \bibnamefont
  {Diddams}}, \bibinfo {author} {\bibfnamefont {J.~C.}\ \bibnamefont
  {Bergquist}}, \bibinfo {author} {\bibfnamefont {S.~R.}\ \bibnamefont
  {Jefferts}}, \ and\ \bibinfo {author} {\bibfnamefont {C.~W.}\ \bibnamefont
  {Oates}},\ }\href {\doibase 10.1126/science.1102330} {\bibfield  {journal}
  {\bibinfo  {journal} {Science}\ }\textbf {\bibinfo {volume} {306}},\ \bibinfo
  {pages} {1318} (\bibinfo {year} {2004})}\BibitemShut {NoStop}%
\bibitem [{\citenamefont {Anderson}\ \emph {et~al.}(1995)\citenamefont
  {Anderson}, \citenamefont {Ensher}, \citenamefont {Matthews}, \citenamefont
  {Wieman},\ and\ \citenamefont {Cornell}}]{anderson-95}%
  \BibitemOpen
  \bibfield  {author} {\bibinfo {author} {\bibfnamefont {M.~H.}\ \bibnamefont
  {Anderson}}, \bibinfo {author} {\bibfnamefont {J.~R.}\ \bibnamefont
  {Ensher}}, \bibinfo {author} {\bibfnamefont {M.~R.}\ \bibnamefont
  {Matthews}}, \bibinfo {author} {\bibfnamefont {C.~E.}\ \bibnamefont
  {Wieman}}, \ and\ \bibinfo {author} {\bibfnamefont {E.~A.}\ \bibnamefont
  {Cornell}},\ }\href {\doibase 10.1126/science.269.5221.198} {\bibfield
  {journal} {\bibinfo  {journal} {Science}\ }\textbf {\bibinfo {volume}
  {269}},\ \bibinfo {pages} {198} (\bibinfo {year} {1995})}\BibitemShut
  {NoStop}%
\bibitem [{\citenamefont {Bradley}\ \emph {et~al.}(1995)\citenamefont
  {Bradley}, \citenamefont {Sackett}, \citenamefont {Tollett},\ and\
  \citenamefont {Hulet}}]{bradley-95}%
  \BibitemOpen
  \bibfield  {author} {\bibinfo {author} {\bibfnamefont {C.~C.}\ \bibnamefont
  {Bradley}}, \bibinfo {author} {\bibfnamefont {C.~A.}\ \bibnamefont
  {Sackett}}, \bibinfo {author} {\bibfnamefont {J.~J.}\ \bibnamefont
  {Tollett}}, \ and\ \bibinfo {author} {\bibfnamefont {R.~G.}\ \bibnamefont
  {Hulet}},\ }\href {\doibase 10.1103/PhysRevLett.75.1687} {\bibfield
  {journal} {\bibinfo  {journal} {Phys. Rev. Lett.}\ }\textbf {\bibinfo
  {volume} {75}},\ \bibinfo {pages} {1687} (\bibinfo {year}
  {1995})}\BibitemShut {NoStop}%
\bibitem [{\citenamefont {Davis}\ \emph {et~al.}(1995)\citenamefont {Davis},
  \citenamefont {Mewes}, \citenamefont {Andrews}, \citenamefont {van Druten},
  \citenamefont {Durfee}, \citenamefont {Kurn},\ and\ \citenamefont
  {Ketterle}}]{davis-95}%
  \BibitemOpen
  \bibfield  {author} {\bibinfo {author} {\bibfnamefont {K.~B.}\ \bibnamefont
  {Davis}}, \bibinfo {author} {\bibfnamefont {M.~O.}\ \bibnamefont {Mewes}},
  \bibinfo {author} {\bibfnamefont {M.~R.}\ \bibnamefont {Andrews}}, \bibinfo
  {author} {\bibfnamefont {N.~J.}\ \bibnamefont {van Druten}}, \bibinfo
  {author} {\bibfnamefont {D.~S.}\ \bibnamefont {Durfee}}, \bibinfo {author}
  {\bibfnamefont {D.~M.}\ \bibnamefont {Kurn}}, \ and\ \bibinfo {author}
  {\bibfnamefont {W.}~\bibnamefont {Ketterle}},\ }\href {\doibase
  10.1103/PhysRevLett.75.3969} {\bibfield  {journal} {\bibinfo  {journal}
  {Phys. Rev. Lett.}\ }\textbf {\bibinfo {volume} {75}},\ \bibinfo {pages}
  {3969} (\bibinfo {year} {1995})}\BibitemShut {NoStop}%
\bibitem [{\citenamefont {Mitroy}\ \emph {et~al.}(2010)\citenamefont {Mitroy},
  \citenamefont {Safronova},\ and\ \citenamefont {Clark}}]{mitroy-10}%
  \BibitemOpen
  \bibfield  {author} {\bibinfo {author} {\bibfnamefont {J.}~\bibnamefont
  {Mitroy}}, \bibinfo {author} {\bibfnamefont {M.~S.}\ \bibnamefont
  {Safronova}}, \ and\ \bibinfo {author} {\bibfnamefont {C.~W.}\ \bibnamefont
  {Clark}},\ }\href {http://stacks.iop.org/0953-4075/43/i=20/a=202001}
  {\bibfield  {journal} {\bibinfo  {journal} {J. Phys. B}\ }\textbf {\bibinfo
  {volume} {43}},\ \bibinfo {pages} {202001} (\bibinfo {year}
  {2010})}\BibitemShut {NoStop}%
\bibitem [{\citenamefont {Coester}(1958)}]{coester-58}%
  \BibitemOpen
  \bibfield  {author} {\bibinfo {author} {\bibfnamefont {F.}~\bibnamefont
  {Coester}},\ }\href {\doibase 10.1016/0029-5582(58)90280-3} {\bibfield
  {journal} {\bibinfo  {journal} {Nucl. Phys.}\ }\textbf {\bibinfo {volume}
  {7}},\ \bibinfo {pages} {421 } (\bibinfo {year} {1958})}\BibitemShut
  {NoStop}%
\bibitem [{\citenamefont {Coester}\ and\ \citenamefont
  {K{\"{u}}mmel}(1960)}]{coester-60}%
  \BibitemOpen
  \bibfield  {author} {\bibinfo {author} {\bibfnamefont {F.}~\bibnamefont
  {Coester}}\ and\ \bibinfo {author} {\bibfnamefont {H.}~\bibnamefont
  {K{\"{u}}mmel}},\ }\href {\doibase 10.1016/0029-5582(60)90140-1} {\bibfield
  {journal} {\bibinfo  {journal} {Nucl. Phys.}\ }\textbf {\bibinfo {volume}
  {17}},\ \bibinfo {pages} {477 } (\bibinfo {year} {1960})}\BibitemShut
  {NoStop}%
\bibitem [{\citenamefont {Bartlett}\ and\ \citenamefont
  {Musia\l{}}(2007)}]{bartlett-07}%
  \BibitemOpen
  \bibfield  {author} {\bibinfo {author} {\bibfnamefont {R.~J.}\ \bibnamefont
  {Bartlett}}\ and\ \bibinfo {author} {\bibfnamefont {M.}~\bibnamefont
  {Musia\l{}}},\ }\href {\doibase 10.1103/RevModPhys.79.291} {\bibfield
  {journal} {\bibinfo  {journal} {Rev. Mod. Phys.}\ }\textbf {\bibinfo {volume}
  {79}},\ \bibinfo {pages} {291} (\bibinfo {year} {2007})}\BibitemShut
  {NoStop}%
\bibitem [{\citenamefont {Mani}\ \emph {et~al.}(2009)\citenamefont {Mani},
  \citenamefont {Latha},\ and\ \citenamefont {Angom}}]{mani-09}%
  \BibitemOpen
  \bibfield  {author} {\bibinfo {author} {\bibfnamefont {B.~K.}\ \bibnamefont
  {Mani}}, \bibinfo {author} {\bibfnamefont {K.~V.~P.}\ \bibnamefont {Latha}},
  \ and\ \bibinfo {author} {\bibfnamefont {D.}~\bibnamefont {Angom}},\ }\href
  {\doibase 10.1103/PhysRevA.80.062505} {\bibfield  {journal} {\bibinfo
  {journal} {Phys. Rev. A}\ }\textbf {\bibinfo {volume} {80}},\ \bibinfo
  {pages} {062505} (\bibinfo {year} {2009})}\BibitemShut {NoStop}%
\bibitem [{\citenamefont {Nataraj}\ \emph {et~al.}(2008)\citenamefont
  {Nataraj}, \citenamefont {Sahoo}, \citenamefont {Das},\ and\ \citenamefont
  {Mukherjee}}]{nataraj-08}%
  \BibitemOpen
  \bibfield  {author} {\bibinfo {author} {\bibfnamefont {H.~S.}\ \bibnamefont
  {Nataraj}}, \bibinfo {author} {\bibfnamefont {B.~K.}\ \bibnamefont {Sahoo}},
  \bibinfo {author} {\bibfnamefont {B.~P.}\ \bibnamefont {Das}}, \ and\
  \bibinfo {author} {\bibfnamefont {D.}~\bibnamefont {Mukherjee}},\ }\href
  {\doibase 10.1103/PhysRevLett.101.033002} {\bibfield  {journal} {\bibinfo
  {journal} {Phys. Rev. Lett.}\ }\textbf {\bibinfo {volume} {101}},\ \bibinfo
  {pages} {033002} (\bibinfo {year} {2008})}\BibitemShut {NoStop}%
\bibitem [{\citenamefont {Pal}\ \emph {et~al.}(2007)\citenamefont {Pal},
  \citenamefont {Safronova}, \citenamefont {Johnson}, \citenamefont
  {Derevianko},\ and\ \citenamefont {Porsev}}]{pal-07}%
  \BibitemOpen
  \bibfield  {author} {\bibinfo {author} {\bibfnamefont {R.}~\bibnamefont
  {Pal}}, \bibinfo {author} {\bibfnamefont {M.~S.}\ \bibnamefont {Safronova}},
  \bibinfo {author} {\bibfnamefont {W.~R.}\ \bibnamefont {Johnson}}, \bibinfo
  {author} {\bibfnamefont {A.}~\bibnamefont {Derevianko}}, \ and\ \bibinfo
  {author} {\bibfnamefont {S.~G.}\ \bibnamefont {Porsev}},\ }\href {\doibase
  10.1103/PhysRevA.75.042515} {\bibfield  {journal} {\bibinfo  {journal} {Phys.
  Rev. A}\ }\textbf {\bibinfo {volume} {75}},\ \bibinfo {pages} {042515}
  (\bibinfo {year} {2007})}\BibitemShut {NoStop}%
\bibitem [{\citenamefont {Gopakumar}\ \emph {et~al.}(2001)\citenamefont
  {Gopakumar}, \citenamefont {Merlitz}, \citenamefont {Majumder}, \citenamefont
  {Chaudhuri}, \citenamefont {Das}, \citenamefont {Mahapatra},\ and\
  \citenamefont {Mukherjee}}]{geetha-01}%
  \BibitemOpen
  \bibfield  {author} {\bibinfo {author} {\bibfnamefont {G.}~\bibnamefont
  {Gopakumar}}, \bibinfo {author} {\bibfnamefont {H.}~\bibnamefont {Merlitz}},
  \bibinfo {author} {\bibfnamefont {S.}~\bibnamefont {Majumder}}, \bibinfo
  {author} {\bibfnamefont {R.~K.}\ \bibnamefont {Chaudhuri}}, \bibinfo {author}
  {\bibfnamefont {B.~P.}\ \bibnamefont {Das}}, \bibinfo {author} {\bibfnamefont
  {U.~S.}\ \bibnamefont {Mahapatra}}, \ and\ \bibinfo {author} {\bibfnamefont
  {D.}~\bibnamefont {Mukherjee}},\ }\href {\doibase 10.1103/PhysRevA.64.032502}
  {\bibfield  {journal} {\bibinfo  {journal} {Phys. Rev. A}\ }\textbf {\bibinfo
  {volume} {64}},\ \bibinfo {pages} {032502} (\bibinfo {year}
  {2001})}\BibitemShut {NoStop}%
\bibitem [{\citenamefont {Isaev}\ \emph {et~al.}(2004)\citenamefont {Isaev},
  \citenamefont {Petrov}, \citenamefont {Mosyagin}, \citenamefont {Titov},
  \citenamefont {Eliav},\ and\ \citenamefont {Kaldor}}]{isaev-04}%
  \BibitemOpen
  \bibfield  {author} {\bibinfo {author} {\bibfnamefont {T.~A.}\ \bibnamefont
  {Isaev}}, \bibinfo {author} {\bibfnamefont {A.~N.}\ \bibnamefont {Petrov}},
  \bibinfo {author} {\bibfnamefont {N.~S.}\ \bibnamefont {Mosyagin}}, \bibinfo
  {author} {\bibfnamefont {A.~V.}\ \bibnamefont {Titov}}, \bibinfo {author}
  {\bibfnamefont {E.}~\bibnamefont {Eliav}}, \ and\ \bibinfo {author}
  {\bibfnamefont {U.}~\bibnamefont {Kaldor}},\ }\href {\doibase
  10.1103/PhysRevA.69.030501} {\bibfield  {journal} {\bibinfo  {journal} {Phys.
  Rev. A}\ }\textbf {\bibinfo {volume} {69}},\ \bibinfo {pages} {030501}
  (\bibinfo {year} {2004})}\BibitemShut {NoStop}%
\bibitem [{\citenamefont {Hagen}\ \emph {et~al.}(2008)\citenamefont {Hagen},
  \citenamefont {Papenbrock}, \citenamefont {Dean},\ and\ \citenamefont
  {Hjorth-Jensen}}]{hagen-08}%
  \BibitemOpen
  \bibfield  {author} {\bibinfo {author} {\bibfnamefont {G.}~\bibnamefont
  {Hagen}}, \bibinfo {author} {\bibfnamefont {T.}~\bibnamefont {Papenbrock}},
  \bibinfo {author} {\bibfnamefont {D.~J.}\ \bibnamefont {Dean}}, \ and\
  \bibinfo {author} {\bibfnamefont {M.}~\bibnamefont {Hjorth-Jensen}},\ }\href
  {\doibase 10.1103/PhysRevLett.101.092502} {\bibfield  {journal} {\bibinfo
  {journal} {Phys. Rev. Lett.}\ }\textbf {\bibinfo {volume} {101}},\ \bibinfo
  {pages} {092502} (\bibinfo {year} {2008})}\BibitemShut {NoStop}%
\bibitem [{\citenamefont {Bishop}\ \emph {et~al.}(2009)\citenamefont {Bishop},
  \citenamefont {Li}, \citenamefont {Farnell},\ and\ \citenamefont
  {Campbell}}]{bishop-09}%
  \BibitemOpen
  \bibfield  {author} {\bibinfo {author} {\bibfnamefont {R.~F.}\ \bibnamefont
  {Bishop}}, \bibinfo {author} {\bibfnamefont {P.~H.~Y.}\ \bibnamefont {Li}},
  \bibinfo {author} {\bibfnamefont {D.~J.~J.}\ \bibnamefont {Farnell}}, \ and\
  \bibinfo {author} {\bibfnamefont {C.~E.}\ \bibnamefont {Campbell}},\ }\href
  {\doibase 10.1103/PhysRevB.79.174405} {\bibfield  {journal} {\bibinfo
  {journal} {Phys. Rev. B}\ }\textbf {\bibinfo {volume} {79}},\ \bibinfo
  {pages} {174405} (\bibinfo {year} {2009})}\BibitemShut {NoStop}%
\bibitem [{\citenamefont {Thakkar}\ \emph {et~al.}(1992)\citenamefont
  {Thakkar}, \citenamefont {Hettema},\ and\ \citenamefont
  {Wormer}}]{thakkar-92}%
  \BibitemOpen
  \bibfield  {author} {\bibinfo {author} {\bibfnamefont {A.~J.}\ \bibnamefont
  {Thakkar}}, \bibinfo {author} {\bibfnamefont {H.}~\bibnamefont {Hettema}}, \
  and\ \bibinfo {author} {\bibfnamefont {P.~E.~S.}\ \bibnamefont {Wormer}},\
  }\href {\doibase 10.1063/1.463012} {\bibfield  {journal} {\bibinfo  {journal}
  {J. Chem. Phys.}\ }\textbf {\bibinfo {volume} {97}},\ \bibinfo {pages} {3252}
  (\bibinfo {year} {1992})}\BibitemShut {NoStop}%
\bibitem [{\citenamefont {Soldan}\ \emph {et~al.}(2001)\citenamefont {Soldan},
  \citenamefont {Lee},\ and\ \citenamefont {Wright}}]{soldan-01}%
  \BibitemOpen
  \bibfield  {author} {\bibinfo {author} {\bibfnamefont {P.}~\bibnamefont
  {Soldan}}, \bibinfo {author} {\bibfnamefont {E.~P.~F.}\ \bibnamefont {Lee}},
  \ and\ \bibinfo {author} {\bibfnamefont {T.~G.}\ \bibnamefont {Wright}},\
  }\href {\doibase 10.1039/B105433N} {\bibfield  {journal} {\bibinfo  {journal}
  {Phys. Chem. Chem. Phys.}\ }\textbf {\bibinfo {volume} {3}},\ \bibinfo
  {pages} {4661} (\bibinfo {year} {2001})}\BibitemShut {NoStop}%
\bibitem [{\citenamefont {Nakajima}\ and\ \citenamefont
  {Hirao}(2001)}]{nakajima-01}%
  \BibitemOpen
  \bibfield  {author} {\bibinfo {author} {\bibfnamefont {T.}~\bibnamefont
  {Nakajima}}\ and\ \bibinfo {author} {\bibfnamefont {K.}~\bibnamefont
  {Hirao}},\ }\href@noop {} {\bibfield  {journal} {\bibinfo  {journal} {Chem.
  Lett.}\ }\textbf {\bibinfo {volume} {30}},\ \bibinfo {pages} {766} (\bibinfo
  {year} {2001})}\BibitemShut {NoStop}%
\bibitem [{\citenamefont {Nakajima}\ and\ \citenamefont
  {Hirao}(2000)}]{nakajima-00}%
  \BibitemOpen
  \bibfield  {author} {\bibinfo {author} {\bibfnamefont {T.}~\bibnamefont
  {Nakajima}}\ and\ \bibinfo {author} {\bibfnamefont {K.}~\bibnamefont
  {Hirao}},\ }\href {\doibase 10.1063/1.1316037} {\bibfield  {journal}
  {\bibinfo  {journal} {J. Chem. Phys.}\ }\textbf {\bibinfo {volume} {113}},\
  \bibinfo {pages} {7786} (\bibinfo {year} {2000})}\BibitemShut {NoStop}%
\bibitem [{\citenamefont {Mohr}\ \emph {et~al.}(2012)\citenamefont {Mohr},
  \citenamefont {Taylor},\ and\ \citenamefont {Newell}}]{codata-10}%
  \BibitemOpen
  \bibfield  {author} {\bibinfo {author} {\bibfnamefont {P.~J.}\ \bibnamefont
  {Mohr}}, \bibinfo {author} {\bibfnamefont {B.~N.}\ \bibnamefont {Taylor}}, \
  and\ \bibinfo {author} {\bibfnamefont {D.~B.}\ \bibnamefont {Newell}},\
  }\href@noop {} {\enquote {\bibinfo {title} {Codata recommended values of the
  fundamental physical constants: 2010},}\ } (\bibinfo {year}
  {2012})\BibitemShut {NoStop}%
\bibitem [{\citenamefont {Brown}\ and\ \citenamefont
  {Ravenhall}(1951)}]{brown-51}%
  \BibitemOpen
  \bibfield  {author} {\bibinfo {author} {\bibfnamefont {G.~E.}\ \bibnamefont
  {Brown}}\ and\ \bibinfo {author} {\bibfnamefont {D.~G.}\ \bibnamefont
  {Ravenhall}},\ }\href {\doibase 10.1098/rspa.1951.0181} {\bibfield  {journal}
  {\bibinfo  {journal} {Proceedings of the Royal Society of London. Series A.
  Mathematical and Physical Sciences}\ }\textbf {\bibinfo {volume} {208}},\
  \bibinfo {pages} {552} (\bibinfo {year} {1951})}\BibitemShut {NoStop}%
\bibitem [{\citenamefont {Sucher}(1980)}]{sucher-80}%
  \BibitemOpen
  \bibfield  {author} {\bibinfo {author} {\bibfnamefont {J.}~\bibnamefont
  {Sucher}},\ }\href {\doibase 10.1103/PhysRevA.22.348} {\bibfield  {journal}
  {\bibinfo  {journal} {Phys. Rev. A}\ }\textbf {\bibinfo {volume} {22}},\
  \bibinfo {pages} {348} (\bibinfo {year} {1980})}\BibitemShut {NoStop}%
\bibitem [{\citenamefont {Mani}\ and\ \citenamefont {Angom}(2011)}]{mani-11-3}%
  \BibitemOpen
  \bibfield  {author} {\bibinfo {author} {\bibfnamefont {B.~K.}\ \bibnamefont
  {Mani}}\ and\ \bibinfo {author} {\bibfnamefont {D.}~\bibnamefont {Angom}},\
  }\href@noop {} {\bibfield  {journal} {\bibinfo  {journal} {ArXiv e-prints}\ }
  (\bibinfo {year} {2011})},\ \Eprint {http://arxiv.org/abs/1105.3447}
  {1105.3447} \BibitemShut {NoStop}%
\bibitem [{\citenamefont {Lindgren}\ and\ \citenamefont
  {Morrison}(1986)}]{lindgren-86}%
  \BibitemOpen
  \bibfield  {author} {\bibinfo {author} {\bibfnamefont {I.}~\bibnamefont
  {Lindgren}}\ and\ \bibinfo {author} {\bibfnamefont {J.}~\bibnamefont
  {Morrison}},\ }\href {http://books.google.co.in/books?id=aQjwAAAAMAAJ} {\emph
  {\bibinfo {title} {Atomic Many-Body Theory}}}\ (\bibinfo  {publisher}
  {Springer},\ \bibinfo {year} {2nd Edition, 1986})\BibitemShut {NoStop}%
\bibitem [{\citenamefont {Mohanty}\ and\ \citenamefont
  {Clementi}(1990)}]{mohanty-90}%
  \BibitemOpen
  \bibfield  {author} {\bibinfo {author} {\bibfnamefont {A.~K.}\ \bibnamefont
  {Mohanty}}\ and\ \bibinfo {author} {\bibfnamefont {E.}~\bibnamefont
  {Clementi}},\ }\href {\doibase 10.1063/1.459110} {\bibfield  {journal}
  {\bibinfo  {journal} {J. Chem. Phys.}\ }\textbf {\bibinfo {volume} {93}},\
  \bibinfo {pages} {1829} (\bibinfo {year} {1990})}\BibitemShut {NoStop}%
\bibitem [{\citenamefont {Stanton}\ and\ \citenamefont
  {Havriliak}(1984)}]{stanton-84}%
  \BibitemOpen
  \bibfield  {author} {\bibinfo {author} {\bibfnamefont {R.~E.}\ \bibnamefont
  {Stanton}}\ and\ \bibinfo {author} {\bibfnamefont {S.}~\bibnamefont
  {Havriliak}},\ }\href {\doibase 10.1063/1.447865} {\bibfield  {journal}
  {\bibinfo  {journal} {J. Chem. Phys.}\ }\textbf {\bibinfo {volume} {81}},\
  \bibinfo {pages} {1910} (\bibinfo {year} {1984})}\BibitemShut {NoStop}%
\bibitem [{\citenamefont {Chaudhuri}\ \emph {et~al.}(1999)\citenamefont
  {Chaudhuri}, \citenamefont {Panda},\ and\ \citenamefont
  {Das}}]{chaudhuri-99}%
  \BibitemOpen
  \bibfield  {author} {\bibinfo {author} {\bibfnamefont {R.~K.}\ \bibnamefont
  {Chaudhuri}}, \bibinfo {author} {\bibfnamefont {P.~K.}\ \bibnamefont
  {Panda}}, \ and\ \bibinfo {author} {\bibfnamefont {B.~P.}\ \bibnamefont
  {Das}},\ }\href {\doibase 10.1103/PhysRevA.59.1187} {\bibfield  {journal}
  {\bibinfo  {journal} {Phys. Rev. A}\ }\textbf {\bibinfo {volume} {59}},\
  \bibinfo {pages} {1187} (\bibinfo {year} {1999})}\BibitemShut {NoStop}%
\bibitem [{\citenamefont {Parpia}\ \emph {et~al.}(1996)\citenamefont {Parpia},
  \citenamefont {Froese~Fischer},\ and\ \citenamefont {Grant}}]{parpia-96}%
  \BibitemOpen
  \bibfield  {author} {\bibinfo {author} {\bibfnamefont {F.~A.}\ \bibnamefont
  {Parpia}}, \bibinfo {author} {\bibfnamefont {C.}~\bibnamefont
  {Froese~Fischer}}, \ and\ \bibinfo {author} {\bibfnamefont {I.~P.}\
  \bibnamefont {Grant}},\ }\href {\doibase 10.1016/0010-4655(95)00136-0}
  {\bibfield  {journal} {\bibinfo  {journal} {Comp. Phys. Comm.}\ }\textbf
  {\bibinfo {volume} {94}},\ \bibinfo {pages} {249 } (\bibinfo {year}
  {1996})}\BibitemShut {NoStop}%
\bibitem [{\citenamefont {Pulay}(1980)}]{pulay-80}%
  \BibitemOpen
  \bibfield  {author} {\bibinfo {author} {\bibfnamefont {P.}~\bibnamefont
  {Pulay}},\ }\href {\doibase 10.1016/0009-2614(80)80396-4} {\bibfield
  {journal} {\bibinfo  {journal} {Chem. Phys. Lett.}\ }\textbf {\bibinfo
  {volume} {73}},\ \bibinfo {pages} {393 } (\bibinfo {year}
  {1980})}\BibitemShut {NoStop}%
\bibitem [{\citenamefont {Mann}\ and\ \citenamefont {Johnson}(1971)}]{mann-71}%
  \BibitemOpen
  \bibfield  {author} {\bibinfo {author} {\bibfnamefont {J.~B.}\ \bibnamefont
  {Mann}}\ and\ \bibinfo {author} {\bibfnamefont {W.~R.}\ \bibnamefont
  {Johnson}},\ }\href {\doibase 10.1103/PhysRevA.4.41} {\bibfield  {journal}
  {\bibinfo  {journal} {Phys. Rev. A}\ }\textbf {\bibinfo {volume} {4}},\
  \bibinfo {pages} {41} (\bibinfo {year} {1971})}\BibitemShut {NoStop}%
\bibitem [{\citenamefont {Grant}\ and\ \citenamefont {Pyper}(1976)}]{grant-76}%
  \BibitemOpen
  \bibfield  {author} {\bibinfo {author} {\bibfnamefont {I.~P.}\ \bibnamefont
  {Grant}}\ and\ \bibinfo {author} {\bibfnamefont {N.~C.}\ \bibnamefont
  {Pyper}},\ }\href {http://stacks.iop.org/0022-3700/9/i=5/a=019} {\bibfield
  {journal} {\bibinfo  {journal} {J. Phys. B}\ }\textbf {\bibinfo {volume}
  {9}},\ \bibinfo {pages} {761} (\bibinfo {year} {1976})}\BibitemShut {NoStop}%
\bibitem [{\citenamefont {Grant}\ and\ \citenamefont
  {McKenzie}(1980)}]{grant-80}%
  \BibitemOpen
  \bibfield  {author} {\bibinfo {author} {\bibfnamefont {I.~P.}\ \bibnamefont
  {Grant}}\ and\ \bibinfo {author} {\bibfnamefont {B.~J.}\ \bibnamefont
  {McKenzie}},\ }\href {http://stacks.iop.org/0022-3700/13/i=14/a=007}
  {\bibfield  {journal} {\bibinfo  {journal} {J. Phys. B}\ }\textbf {\bibinfo
  {volume} {13}},\ \bibinfo {pages} {2671} (\bibinfo {year}
  {1980})}\BibitemShut {NoStop}%
\bibitem [{\citenamefont {Quiney}(2003)}]{quiney-03}%
  \BibitemOpen
  \bibfield  {author} {\bibinfo {author} {\bibfnamefont {H.~M.}\ \bibnamefont
  {Quiney}},\ }in\ \href@noop {} {\emph {\bibinfo {booktitle} {Handbook of
  Molecular Physics and Quantum Chemistry}}},\ Vol.~\bibinfo {volume} {2},\
  \bibinfo {editor} {edited by\ \bibinfo {editor} {\bibfnamefont
  {S.}~\bibnamefont {Wilson}}, \bibinfo {editor} {\bibfnamefont {P.~P.}\
  \bibnamefont {Bernath}}, \ and\ \bibinfo {editor} {\bibfnamefont
  {R.}~\bibnamefont {McWeeny}}}\ (\bibinfo  {publisher} {John Wiley \& Sons
  Ltd., Chichester},\ \bibinfo {year} {2003})\ pp.\ \bibinfo {pages}
  {444--483}\BibitemShut {NoStop}%
\bibitem [{\citenamefont {Clementi}(1991)}]{clementi-91}%
  \BibitemOpen
  \bibfield  {author} {\bibinfo {author} {\bibfnamefont {E.}~\bibnamefont
  {Clementi}},\ }\href {http://books.google.co.in/books?id=vZlR9sXdGowC} {\emph
  {\bibinfo {title} {Modern Techniques in Computational Chemistry:
  MOTECC-91}}}\ (\bibinfo  {publisher} {ESCOM},\ \bibinfo {year}
  {1991})\BibitemShut {NoStop}%
\bibitem [{\citenamefont {Parpia}\ \emph {et~al.}(1992)\citenamefont {Parpia},
  \citenamefont {Mohanty},\ and\ \citenamefont {Clementi}}]{parpia-92}%
  \BibitemOpen
  \bibfield  {author} {\bibinfo {author} {\bibfnamefont {F.~A.}\ \bibnamefont
  {Parpia}}, \bibinfo {author} {\bibfnamefont {A.~K.}\ \bibnamefont {Mohanty}},
  \ and\ \bibinfo {author} {\bibfnamefont {E.}~\bibnamefont {Clementi}},\
  }\href {http://stacks.iop.org/0953-4075/25/i=1/a=007} {\bibfield  {journal}
  {\bibinfo  {journal} {J. Phys. B}\ }\textbf {\bibinfo {volume} {25}},\
  \bibinfo {pages} {1} (\bibinfo {year} {1992})}\BibitemShut {NoStop}%
\bibitem [{\citenamefont {Ishikawa}\ \emph {et~al.}(1991)\citenamefont
  {Ishikawa}, \citenamefont {Quiney},\ and\ \citenamefont
  {Malli}}]{ishikawa-91}%
  \BibitemOpen
  \bibfield  {author} {\bibinfo {author} {\bibfnamefont {Y.}~\bibnamefont
  {Ishikawa}}, \bibinfo {author} {\bibfnamefont {H.~M.}\ \bibnamefont
  {Quiney}}, \ and\ \bibinfo {author} {\bibfnamefont {G.~L.}\ \bibnamefont
  {Malli}},\ }\href {\doibase 10.1103/PhysRevA.43.3270} {\bibfield  {journal}
  {\bibinfo  {journal} {Phys. Rev. A}\ }\textbf {\bibinfo {volume} {43}},\
  \bibinfo {pages} {3270} (\bibinfo {year} {1991})}\BibitemShut {NoStop}%
\bibitem [{\citenamefont {Langhoff}\ and\ \citenamefont
  {Karplus}(1969)}]{langhoff-69}%
  \BibitemOpen
  \bibfield  {author} {\bibinfo {author} {\bibfnamefont {P.~W.}\ \bibnamefont
  {Langhoff}}\ and\ \bibinfo {author} {\bibfnamefont {M.}~\bibnamefont
  {Karplus}},\ }\href {\doibase 10.1364/JOSA.59.000863} {\bibfield  {journal}
  {\bibinfo  {journal} {J. Opt. Soc. Am.}\ }\textbf {\bibinfo {volume} {59}},\
  \bibinfo {pages} {863} (\bibinfo {year} {1969})}\BibitemShut {NoStop}%
\bibitem [{\citenamefont {Orcutt}\ and\ \citenamefont
  {Cole}(1967)}]{orcutt-67}%
  \BibitemOpen
  \bibfield  {author} {\bibinfo {author} {\bibfnamefont {R.~H.}\ \bibnamefont
  {Orcutt}}\ and\ \bibinfo {author} {\bibfnamefont {R.~H.}\ \bibnamefont
  {Cole}},\ }\href {\doibase 10.1063/1.1840728} {\bibfield  {journal} {\bibinfo
   {journal} {J. Chem. Phys.}\ }\textbf {\bibinfo {volume} {46}},\ \bibinfo
  {pages} {697} (\bibinfo {year} {1967})}\BibitemShut {NoStop}%
\bibitem [{\citenamefont {Chattopadhyay}\ \emph {et~al.}(2012)\citenamefont
  {Chattopadhyay}, \citenamefont {Mani},\ and\ \citenamefont
  {Angom}}]{chattopadhyay-12}%
  \BibitemOpen
  \bibfield  {author} {\bibinfo {author} {\bibfnamefont {S.}~\bibnamefont
  {Chattopadhyay}}, \bibinfo {author} {\bibfnamefont {B.~K.}\ \bibnamefont
  {Mani}}, \ and\ \bibinfo {author} {\bibfnamefont {D.}~\bibnamefont {Angom}},\
  }\href {\doibase 10.1103/PhysRevA.86.022522} {\bibfield  {journal} {\bibinfo
  {journal} {Phys. Rev. A}\ }\textbf {\bibinfo {volume} {86}},\ \bibinfo
  {pages} {022522} (\bibinfo {year} {2012})}\BibitemShut {NoStop}%
\bibitem [{\citenamefont {Mani}\ and\ \citenamefont {Angom}(2010)}]{mani-10}%
  \BibitemOpen
  \bibfield  {author} {\bibinfo {author} {\bibfnamefont {B.~K.}\ \bibnamefont
  {Mani}}\ and\ \bibinfo {author} {\bibfnamefont {D.}~\bibnamefont {Angom}},\
  }\href {\doibase 10.1103/PhysRevA.81.042514} {\bibfield  {journal} {\bibinfo
  {journal} {Phys. Rev. A}\ }\textbf {\bibinfo {volume} {81}},\ \bibinfo
  {pages} {042514} (\bibinfo {year} {2010})}\BibitemShut {NoStop}%
\end{thebibliography}%
\bibliographystyle{apsrev4-1}

\end{document}